\documentclass[prb,superscriptaddress,twocolumn,amsmath,amssymb]{revtex4-2}

\usepackage{graphicx}
\usepackage{dcolumn}
\usepackage{bm}
\usepackage{mathrsfs}
\usepackage{subfigure}
\usepackage{natbib}
\usepackage{amsmath}
\usepackage{amssymb}
\usepackage{Jonasmacros}
\usepackage{mathptmx}
\usepackage{txfonts}
\usepackage{pifont}

\usepackage[colorlinks,citecolor=blue,linkcolor=Fuchsia]{hyperref}
\usepackage[usenames,dvipsnames,svgnames]{xcolor}

\begin{document}
\date{\today} 

\title{The Chiral Induced Spin Selectivity Effect\\What It Is, What It Is Not, And Why It Matters}

\author{J. Fransson}
\email{Jonas.Fransson@physics.uu.se}
\affiliation{Department of Physics and Astronomy, Box 516, 751 21, Uppsala University, Uppsala, Sweden}




\begin{abstract}
The chiral induced spin selectivity effect is an excited states phenomenon, which can be probed using photo-spectroscopy as well as transport measurements. On the one hand such measurements represent averaged quantities, on the other hand nearly all theoretical descriptions, with only a few exceptions, have been concerned with energy dependent properties of the pertinent structures. While those properties may or may not be relevant for the chiral induced spin selectivity effect, many of those properties have been attributed as being the, or part of the, origins of the effect. Here, it is demonstrated that, for instance, the spin-resolved transmission provides little, if any, information about the chiral induced spin selectivity effect. Moreover, although effective single-electron theory can be used in this context, reasons are given for why such descriptions are not viable.
\end{abstract}

\maketitle


\section{Introduction}
\label{sec-introduction}

Chirality is an intrinsic property of some molecules and materials, in which there is no center of inversion or mirror plane. It turns out that, when coupling such structures to the environment, for instance, spin selective processes is an emergent phenomenon \cite{Science.283.814,Science.331.894,APLMaterials.9.040902} which over the past two decades has caused an intensive debate in chemistry and physics communities about its origin.

The chiral induced spin selectivity effect has been reproducibly observed in various measurements, for instance, light exposure \cite{Science.283.814,Science.331.894,PNAS.110.14872,NanoLett.14.6042,NatComms.7.10744,AdvMat.30.1707390,JPhysChemLett.9.2025,JPhysChemC.125.9875,Chirality.33.93}, local probing techniques \cite{NanoLett.11.4652,JPhysChemLett.11.1550,AdvMater.28.1957,NanoLett.19.5167,ACSNano.14.16624}, transport \cite{NatComms.4.2256,JPhysChemLett.10.1139,JPhysChemLett.11.1550} and different types of Hall measurements \cite{NatComms.7.10744,NatComms.8.14567,AdvMat.30.1707390,PhysRevLett.124.166602,PhysRevLett.127.126602}. However, although all measurements entail non-equilibrium conditions, with only a few exceptions, essentially all theoretical accounts of the effect are based on the transmission properties of chiral molecules embedded in a given environment, see for instance Refs. \citenum{JChemPhys.131.014707,EPL.99.17006,JPCM.26.015008,PhysRevB.88.165409,JChemPhys.142.194308,PhysRevE.98.052221,PhysRevB.99.024418,NJP.20.043055,JPhysChemC.123.17043,PhysRevB.85.081404(R),PhysRevLett.108.218102,PNAS.11.11658,JPhysChemC.117.13730,PhysRevB.93.075407,PhysRevB.93.155436,ChemPhys.477.61,JPhysChemLett.9.5453,JPhysChemLett.9.5753,JChemTheoryComput.16.2914,CommunPhys.3.178,NewJPhys.22.113023,PhysRevB.102.035431,NanoLett.21.6696,JPhysChemLett.12.10262,NanoLett.21.10423}.
%
It should be mentioned that despite the concept of the transmission is not a linear response quantity in itself it is, nevertheless, in this context nearly without exception considered in linear response theory and, hence, account for only the ground state properties of the molecule. Moreover, it is also often typically the result of a single particle description which under stationary condition cannot account for the excited states properties that underlie spin selectivity in chiral molecules. Finally, although the intrinsic spin-dependent properties and spin currents of chiral molecules may be interesting quantities in their own rights, see for instance Refs. \citenum{JChemPhys.131.014707,EPL.99.17006,JPCM.26.015008,PhysRevB.88.165409,JChemPhys.142.194308,PhysRevE.98.052221,PhysRevB.99.024418,NJP.20.043055,JPhysChemC.123.17043,PhysRevB.85.081404(R),PhysRevLett.108.218102,PNAS.11.11658,JPhysChemC.117.13730,PhysRevB.93.075407,PhysRevB.93.155436,ChemPhys.477.61,JPhysChemLett.9.5453,JPhysChemLett.9.5753,JChemTheoryComput.16.2914,CommunPhys.3.178,NewJPhys.22.113023,PhysRevB.102.035431,NanoLett.21.6696,JPhysChemLett.12.10262,NanoLett.21.10423,JChemPhys.154.110901,NanoLett.21.8190,PhysRevB.104.024430}, the chiral induced spin selectivity may have little, if anything, to do with those quantities. 

\begin{figure}[t]
\begin{center}
\includegraphics[width=\columnwidth]{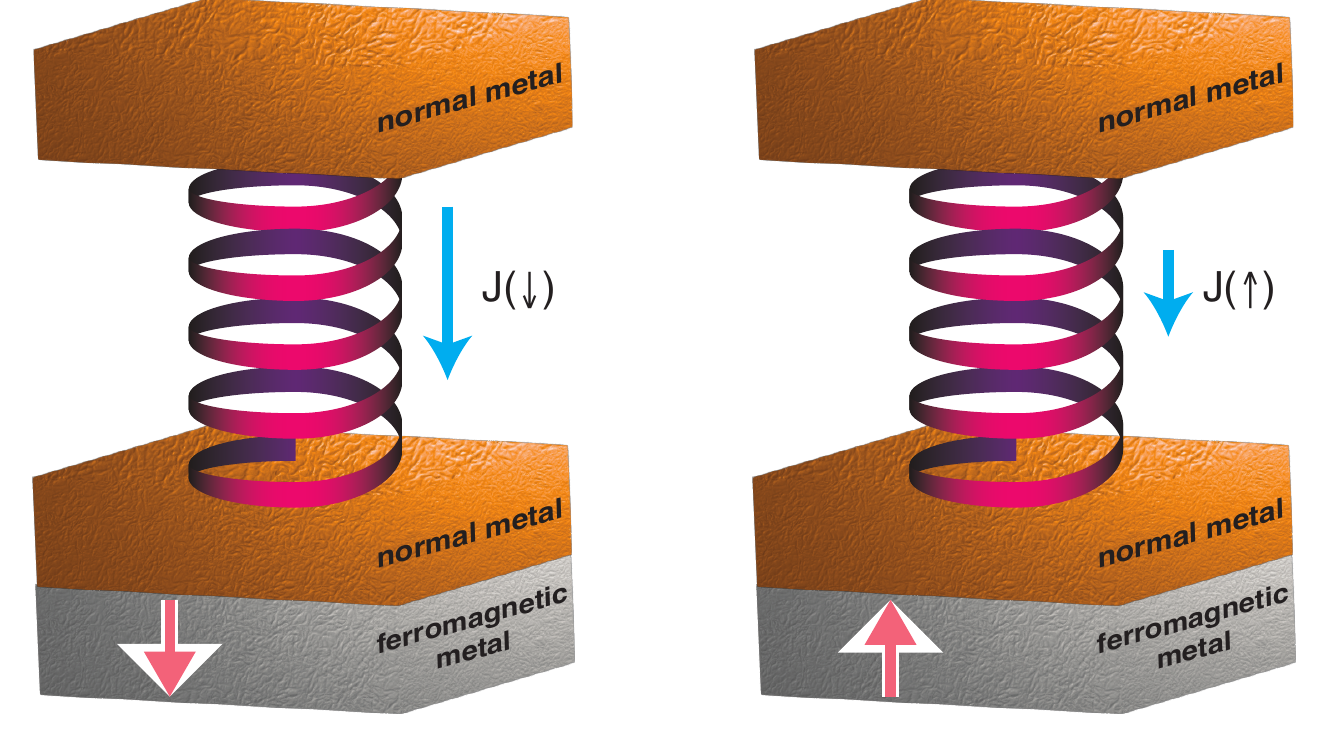}
\end{center}
\caption{Schematic illustration of the transport set-up where the charge current $J$ is measured for magnetic moment of the lower lead oriented in different directions.}
\label{fig-schematic}
\end{figure}

The measurements where the chiral induced spin selectivity is observed is related to a particle flux through the molecule under two different, typically opposite, magnetic conditions. The transport set-ups, in which the charge current is under scrutiny, are intuitive in the sense that there is, at least, one electrode which can be magnetized in different directions, see Figure \ref{fig-schematic}, which thereby enables a control of the external magnetic conditions. Theoretical models to address the measured currents under such conditions have been proposed 
\cite{NanoLett.19.5253,JPhysChemLett.10.7126,NanoLett.20.7077,PhysRevB.102.235416,PhysRevB.102.214303,NanoLett.21.3026,JPhysChemC.125.9875,JACS.143.14235,JPhysChemC.125.23364,JPhysChemLett.13.808,arXiv.2111.12917}, where essentially all these various models comprise a coupling between the electronic degrees of freedom to some other, both Fermionic \cite{JPhysChemLett.10.7126,NanoLett.20.7077,JACS.143.14235,arXiv.2111.12917} and Bosonic \cite{PhysRevB.102.235416,PhysRevB.102.214303,NanoLett.21.3026,JPhysChemC.125.9875,JPhysChemC.125.23364,JPhysChemLett.13.808}, degrees of freedom. Also in the photospectroscopy experiments, a magnetic metal is often coupled to the sample \cite{JPhysChemC.125.9875} and, hence, provide a control of the magnetic conditions in a similar way. The magnetic response can also be controlled using circularly polarized light \cite{Science.283.814,Science.331.894}. However, regardless of which particle flux is investigated, this flux reflects the total charge and magnetic properties of the chiral molecules, but does not provide any detailed information about its spin properties. In words pertaining to transport, the measurement concerns the charge current and nothing else.

The purpose with this article is to first discuss the theoretical framework which is relevant for the transport measurements of the chiral induced spin selectivity effect, and how this may be represented in mathematical form. Here, it is stressed that studying the transmission of the junction is meaningless in itself, unless it is considered as function of the external magnetic environment. It is, furthermore, shown that although the transmission does carry information about the transport properties, this information is difficult to interpret without detailed knowledge about the corresponding actual transport characteristics. Second, it is shown that the chiral induced spin selectivity can be modelled using an effective single electron description, however, such a description can only be used as a guidance for fitting since it cannot say anything about the microscopic origin of the effect.


\section{Results}
\label{sec-results}


\subsection{Transmission and Current}

\begin{figure}[t]
\begin{center}
\includegraphics[width=\columnwidth]{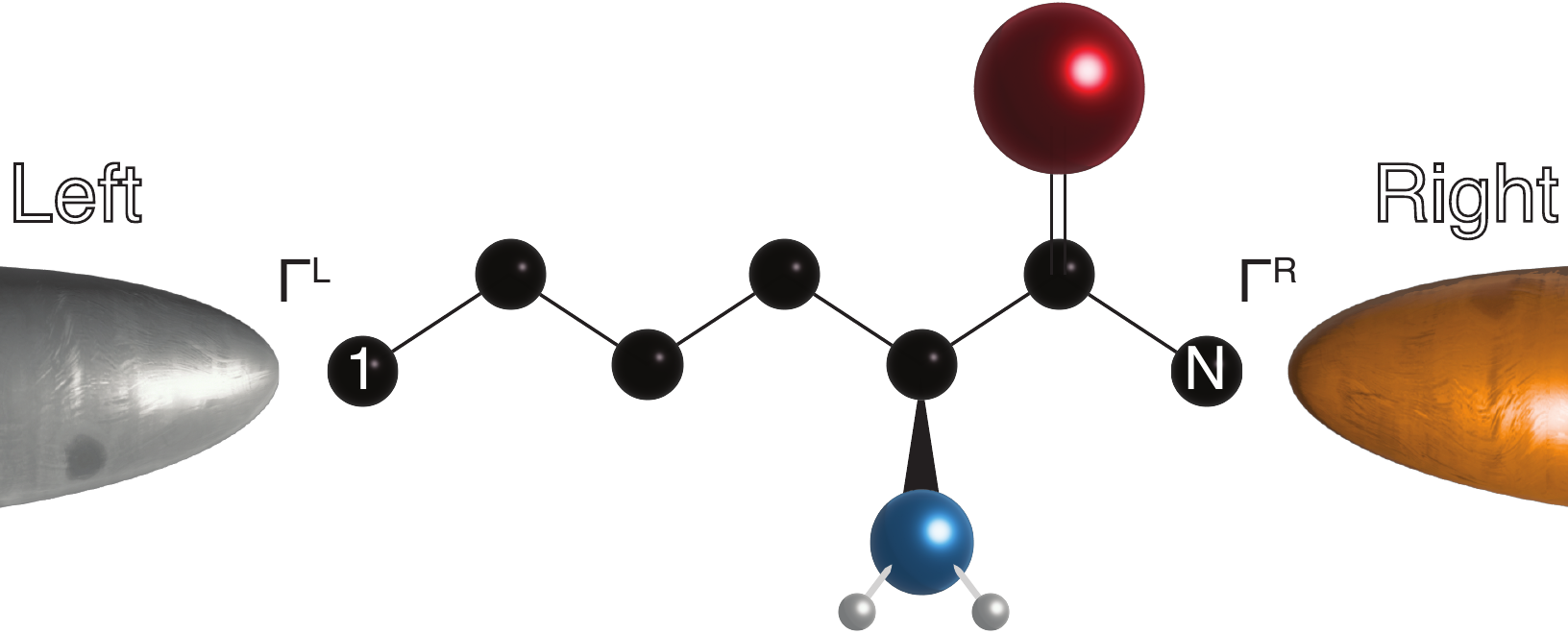}
\end{center}
\caption{Schematic illustration of a chiral molecule in the junction between a ferromagnetic and normal metal.}
\label{fig-MoleculeJunction}
\end{figure}

The first objective is to address the charge current and corresponding transmission through a molecular junction. To this end, consider a generic transport set-up, as schematically illustrated in Figure \ref{fig-MoleculeJunction}, where a chiral molecule is mounted in the junction between a ferromagnetic and normal metallic lead. For the purpose of being concrete, the left lead is assumed to be ferromagnetic whereas the right is a normal metal. In the molecule, the nucleus adjacent to the left (right) lead is labelled 1 ($N$). All discussion henceforth is conducted under the assumption of stationary conditions, such that the charge current through the molecule can be written \cite{PhysRevB.72.075314}
\begin{align}
J=&
	\frac{ie}{h}
	{\rm sp}
	\int
		\bfGamma^L
		\Bigl(
			f_L(\omega)\bfG^>_1(\omega)
			+
			f_L(-\omega)\bfG^<_1(\omega)
		\Bigr)
	d\omega
	.
\label{eq-J}
\end{align}
Here, $\bfG_1^{</>}(\omega)$ denotes the lesser/greater Green function for the site 1 which is directly coupled to the left lead with coupling strength $\bfGamma^L$, whereas $f_L(\omega)$ is the Fermi-Dirac distribution function defined at the electro-chemical potential $\mu_L$ for the left lead. It should be noticed that $\bfG^{</>}_1$ and $\bfGamma^L$ are a $2\times2$ matrices, and that the trace ${\rm sp}$ runs over spin 1/2 space. Also, $e$ denotes the electron charge, $h$ is Planck's constant. Hence, the formula in Eq. \eqref{eq-J} describes the charge current flowing across the interface between the left lead and first site in the molecule. The stationary conditions ensure current conservation such that this charge current is the same as the charge current anywhere else in the system. The non-equilibrium properties at the site 1 are given in terms of its densities of occupied ($\bfG_1^<$) and unoccied ($\bfG_1^>$) electron states which, however, depend on the properties of total system comprising the molecule and metallic leads. This is reflected through the relation $\bfG^{</>}_1=\bfG^r_{1m}\bfSigma^{</>}_{mn}\bfG^a_{n1}$, where the self-energy $\bfSigma_{mn}^{</>}$ accounts for interactions between electrons within the molecule but also with the local environment. In the absence of internal many-body interactions, the self-energy reduces to
\begin{align}
\bfSigma^{</>}_{mn}(\omega)=&
	(\pm i)\delta_{mn}
	\Bigl(
		\delta_{m1}f_L(\pm\omega)\bfGamma^L
		+
		\delta_{mN}f_R(\pm\omega)\bfGamma^R
	\Bigr)
	,
\end{align}
which accounts for the coupling strengths $\bfGamma^\chi$, $\chi=L,R$, between the left ($L$) and right ($R$) leads and sites 1 and $N$, respectively. With these assumptions, the current can be written \cite{PhysRevB.50.5528}
\begin{align}
J=&
	\frac{e}{h}
	\int
		\Bigl(
			f_L(\omega)
			-
			f_R(\omega)
		\Bigr)
		{\rm sp}~
		\calT(\omega)
	d\omega
	,
\end{align}
where $\calT(\omega)=\bfGamma^L\bfG^r_{1N}(\omega)\bfGamma^R\bfG^a_{N1}(\omega)$ can be interpreted as the transmission matrix in the Landauer sense.

By symmetry, the transmission $\calT$ can always be partitioned according to $\calT=\calT_0\sigma^0+\calT_1\cdot\bfsigma$, where $\sigma^0$ and $\bfsigma$ are the identity and vector of Pauli matrices, respectively. Effecting the trace for the $2\times2$-matrix $\calT$ such that the respective charge (${\cal T}_0$) and spin (${\cal T}_1$) components can be defined by ${\cal T}_0={\rm sp}{\cal T}/2$ and ${\cal T}_1={\rm sp}\bfsigma{\cal T}/2$.
In the present discussion, the transmissions can be explicitly obtained in terms of the electronic structure by putting $\bfA=\bfG_{1N}^r$ and $\bfB=\bfG_{N1}^a$, hence, $\bfB^\dagger=\bfA$. It is, furthermore, assumed that the left lead is ferromagnetic, $\bfGamma^L=\Gamma_0(\sigma^0+\bfp_L\cdot\bfsigma)/4$, where the vector $\bfp_L$ parametrizes the orientation of the magnetic moment in the left lead such that $p_L=|\bfp_L|\leq1$, whereas the right lead is a non-magnetic metal, $\bfGamma^R=\Gamma_0\sigma^0/4$. It is convenient to decompose the transmissions into $\calT_m=\sum_{n=0,1}\calT_{mn}$, $m=0,1$, with
\begin{subequations}
\label{eq-T}
\begin{align}
\calT_{00}=&
	\biggl(\frac{\Gamma_0}{4}\biggr)^2
		(
			A_0B_0+\bfA_1\cdot\bfB_1
		)
	,
\label{eq-T00}
\\
\calT_{01}=&
	\biggl(\frac{\Gamma_0}{4}\biggr)^2
		\bfp_L\cdot
		(
			A_0\bfB_1+\bfA_1B_0+i\bfA_1\times\bfB_1
		)
	,
\label{eq-T01}
\\
\calT_{10}=&
	\biggl(\frac{\Gamma_0}{4}\biggr)^2
		(
			A_0\bfB_1+\bfA_1B_0+i\bfA_1\times\bfB_1
		)
	,
\label{eq-T10}
\\
\calT_{11}=&
	\biggl(\frac{\Gamma_0}{4}\biggr)^2
		\Bigl(
			\bfp_L(A_0B_0+\bfA_1\cdot\bfB_1)
\nonumber\\&
			+
			i\bfp_L\times
			(
				A_0\bfB_1+\bfA_1B_0+i\bfA_1\times\bfB_1
			)
		\Bigr)
	,
\label{eq-T11}
\end{align}
\end{subequations}
where $A_0$ ($B_0$) and $\bfA_1$ ($\bfB_1$) correspond to the charge and spin components of $\bfA$ ($\bfB$), such that $\bfA=A_0\sigma^0+\bfA_1\cdot\bfsigma$ ($\bfB=B_0\sigma^0+\bfB_1\cdot\bfsigma$).

Since ${\rm sp}\calT=2\calT_0$ and ${\rm sp}\bfsigma\calT=2\calT_1$, it is clear that only the components $\calT_{0m}$ contribute to the total charge current $J\sim\int(f_L-f_R){\rm sp}\calT d\omega$, whereas the components $\calT_{1m}$ contribute to the spin current $\bfJ_s\sim\int(f_L-f_R){\rm sp}\bfsigma\calT d\omega$. This statement can be understood by considering the spin current ${\bfJ_s}=\partial_t\av{\bfM}=\partial_t\av{\psi^\dagger\bfsigma\psi}$ in analogy with the charge current $J=-e\partial_t\av{N}=-e\partial_t\av{\psi^\dagger\psi}$, where $\psi$ and $\psi^\dagger$ define the annihilation and creation field spinor operators for the total electronic structure, whereas $N$ and $\bfM$ denote the corresponding charge and magnetic density operators, respectively.
It can also be observed that while the spin components of the Green function and coupling matrices are explicitly included in the transmission $\calT_0$, the total transmission included in the charge current is detached from a description in terms of a spin resolved formulation. This latter statement can be understood since the spin-resolved transmission $\calT_{\sigma\sigma'}$, where $\sigma,\sigma'\in\{\up,\down\}$, can be identified as
\begin{align}
\calT_{\sigma\sigma'}=&
	\calT_0\delta_{\sigma\sigma'}
	+
	\calT_1\cdot\bfsigma_{\sigma\sigma'}
	,
\label{eq-Tspinres}
\end{align}
comprising the component $\calT_1$. Hence, despite the apparent spin-dependence of the transmission provided by the second contribution, only the first contribution is relevant for the charge current. Hence, any spin-dependent change in the transmission that can only be traced back to a corresponding change in $\calT_1$, while $\calT_0$ remains unchanged, does not influence the charge current. Through the definition in Eq. \eqref{eq-Tspinres} it can be seen, for instance, that $\calT_{\up\up}$ and $\calT_{\down\down}$ differ by
\begin{align}
\calT_{\up\up}-\calT_{\down\down}=&
	2
	(
		\calT_{10}^z+\calT_{11}^z
	)
	,
\end{align}
that is, this difference does not contribute to the charge current. The spin-resolved transmission is defined without reference to the external conditions $\bfp_L$ and reflects the intrinsic spin properties of the molecule. Since, the chiral induced spin selectivity effect is measured as a response to the external conditions, as discussed above, the appropriate quantity to study is the \emph{magneto-current}, that is, the difference
\begin{align}
\Delta J\equiv&
	J(\up)-J(\down)
	,
\end{align}
where $J(\up)$ ($J(\down)$) denotes the charge current for $\bfp_L$ ($-\bfp_L$), see Figure \ref{fig-schematic}. From this definition one can deduce
\begin{align}
\Delta J=&
	\frac{2e}{h}\int(f_L-f_R)\Bigl(\calT_0(\up)-\calT_0(\down)\Bigr)d\omega
	,
\end{align}
such that the effect can be studied in terms of the transmission $\calT_0(\sigma)$, as function of the orientation of $\bfp_L$. Therefore, this result very clearly illustrates that a non-vanishing spin-resolved transmission $\calT_{\sigma\sigma'}$, in general, and the difference $\calT_{\up\up}-\calT_{\down\down}$, in particular, may have very little, if anything, to do with the comprehension of the chiral induced spin selectivity. It is, then, easy to conclude that the spin-resolved transmission does not represent an appropriate quantity of reference in the context of chiral induced spin selectivity. It, furthermore, means that the origin of the effect is not necessarily related to a spin-resolved transmission coefficient. While this analysis may seem obvious and trivial, investigations of the spin-resolved transmissions have, nevertheless, been the main focus in, for instance, Refs. \citenum{JChemPhys.131.014707,EPL.99.17006,JPCM.26.015008,PhysRevB.88.165409,JChemPhys.142.194308,PhysRevE.98.052221,PhysRevB.99.024418,NJP.20.043055,JPhysChemC.123.17043,PhysRevB.85.081404(R),PhysRevLett.108.218102,PNAS.11.11658,JPhysChemC.117.13730,PhysRevB.93.075407,PhysRevB.93.155436,ChemPhys.477.61,JPhysChemLett.9.5453,JPhysChemLett.9.5753,JChemTheoryComput.16.2914,CommunPhys.3.178,NewJPhys.22.113023,PhysRevB.102.035431,NanoLett.21.6696,JPhysChemLett.12.10262,NanoLett.21.10423}. 


\subsection{Spin-resolved transmission}
The phenomenology of chiral induced spin selectivity can be investigated using a simple effective model for a molecule embedded between two metallic leads. The model comprises a single electronic level, which represents, without loss of generality, the HOMO level under the assumption that the spacing to both HOMO$-n$, $n=1,2,3,\dots$, and LUMO$+n$, $n=0,1,2,\ldots$, is large enough for these levels to not take part in the conduction. The energy spectrum of this level, relative to the equilibrium chemical potential $\mu=0$ is represented by the matrix $\bfepsilon=\dote{0}\sigma^0+\bfepsilon_1\cdot\bfsigma$, where $\dote{0}$ and $\bfepsilon_1$ denote the average energy and local spin anisotropy. The latter energy, $\bfepsilon_1$, represents local Zeeman splitting and spin anisotropy that may be present because of spin-orbit and many-body interactions, but also because of dissipation and coupling to external environment. However, the specific nature of the origins of this anisotropy is not defined here.

In the spirit of the approximation used already previously, the model for this considered configuration can be written
\begin{align}
\Hamil=&
	\sum_{\bfk\in L,R}\psi^\dagger_\bfk\bfepsilon_\bfk\psi_\bfk
	+
	\psi^\dagger\bfepsilon\psi
	+
	\sum_{\bfk\in L,R}\Bigl(\psi_\bfk^\dagger\bfv_\bfk\psi+H.c.\Bigr)
	,
\end{align}
where $\psi^\dagger$ ($\psi$) is the creation (annihilation) spinor for electrons in the molecule (second term), whereas $\psi^\dagger_\bfk$ ($\psi_\bfk$) is the analogous operator for electrons in the left ($L$) and right ($R$) lead defined by the spectrum $\bfepsilon_\bfk=\dote{\bfk}^{(0)}\sigma^0+\bfepsilon_\bfk^{(1)}\cdot\bfsigma$ (first term). The molecule and leads are connected through hybridization (last term) with rate matrix $\bfv_\bfk=v_\bfk^{(0)}\sigma^0+\bfv_\bfk^{(1)}\cdot\bfsigma$, such that the coupling $\bfGamma^\chi$, $\chi=L,R$, can be defined in terms of the electronic density $\bfrho_\bfk$ in the lead $\chi$ through $\bfGamma^\chi=\sum_{\bfk\in\chi}\bfv^\dagger_\bfk\bfrho_\bfk\bfv_\bfk$. Under the conditions introduced previously, $\Gamma^L=\Gamma_0(\sigma^0+\bfp_L\cdot\bfsigma)/4$ and $\bfGamma^R=\Gamma_0\sigma^0/4$.

For this example, the Green function for the molecular level can be written
\begin{align}
\bfG^r(\omega)=&
	\frac{(\omega-\dote{0}+i\Gamma_0/4)\sigma^0+(\bfepsilon_1-i\bfp_L\Gamma_0/8)\cdot\bfsigma}
	{(\omega-\dote{0}+i\Gamma_0/4)^2-(\bfepsilon_1-i\bfp_L\Gamma_0/8)^2}
	.
\label{eq-Gr}
\end{align}
Partitioning into charge, $G_0={\rm sp}\bfG/2$, and spin, $\bfG_1={\rm sp}\bfsigma\bfG/2$, components leads to
\begin{subequations}
\label{eq-G01r}
\begin{align}
G_0^r(\omega)=&
	\frac{\omega-\dote{0}+i\Gamma_0/4}
	{(\omega-\dote{0}+i\Gamma_0/4)^2-(\bfepsilon_1-i\bfp_L\Gamma_0/8)^2}
	,
\label{eq-G0r}
\\
\bfG^r_1(\omega)=&
	\frac{\bfepsilon_1-i\bfp_L\Gamma_0/8}
	{(\omega-\dote{0}+i\Gamma_0/4)^2-(\bfepsilon_1-i\bfp_L\Gamma_0/8)^2}
	.
\label{eq-G1r}
\end{align}
\end{subequations}
In relation to the transmission in the previous subsection, one can define $A_0=G_0^r$ ($B_0=G_0^a$), and $\bfA_1=\bfG^r_1$ ($\bfB_1=\bfG^a_1$), such that $\bfA=\bfG^r$ ($\bfB=\bfG^a$). 

\begin{figure}[t]
\begin{center}
\includegraphics[width=\columnwidth]{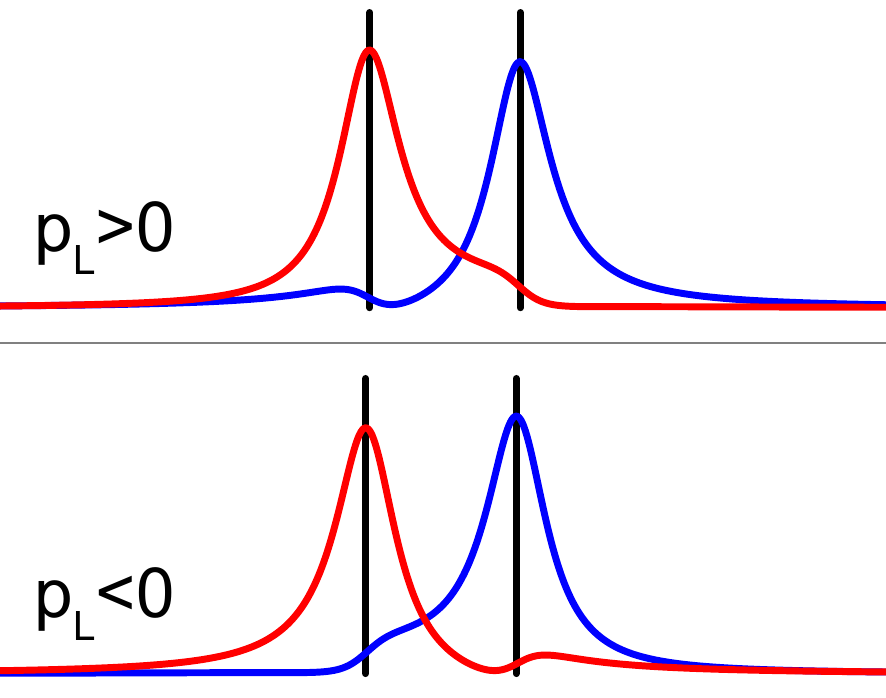}
\end{center}
\caption{Example of the molecular electronic structure for two different, opposite, spin-polizaritions $\bfp_L$ in the left lead, indicating the interchanging level broadenings upon switching sign of $\bfp_L$. Notice also, the slight shifts of the energy levels, indicated with faint lines.}
\label{fig-interchange}
\end{figure}

In these expressions, the poles $E_\pm$ (zeros of the numerator; $(\omega-\dote{0}+i\Gamma_0/4)^2-(\bfepsilon_1-i\bfp_L\Gamma_0/8)^2$) can be written
\begin{subequations}
\begin{align}
E_\pm=&
	\dote{\pm}-\frac{i}{\tau_\pm}
	,
\\
\dote{\pm}=&
	\dote{0}\pm\calR\cos\frac{\varphi}{2}
	,&
\frac{1}{\tau_\pm}=
	\frac{\Gamma_0}{4}
	\pm
	\calR\sin\frac{\varphi}{2}
	,
\end{align}
\end{subequations}
where $\calR^2=[\dote{1}^2-p_L^2(\Gamma_0/8)^2]^2+(\bfepsilon_1\cdot\bfp_L)^2(\Gamma_0/4)^2$ and $\tan\varphi=\bfepsilon_1\cdot\bfp_L\Gamma_0/4[\dote{1}^2-p_L^2(\Gamma_0/8)^2]$. Then, since $\calR$ remains invariant under $\bfp_L\rightarrow-\bfp_L$, while the phase $\varphi$ is odd, the symmetry rules for the poles can be summarized as
\begin{subequations}
\label{eq-Esymmetries}
\begin{align}
\dote{\pm}(-\bfp_L)=&
	\re E_\pm(-\bfp_L)=
	\re E_\pm(\bfp_L)=
	\dote{\pm}(\bfp_L)
	,
\\
\frac{1}{\tau_\pm(-\bfp_L)}=&
	\im E_\pm(-\bfp_L)=
	\im E_\mp(\bfp_L)=
	\frac{1}{\tau_\mp(\bfp_L)}
	.
\end{align}
\end{subequations}
Hence, the level broadenings $1/\tau_\pm$ of the two levels are interchanged upon switching the sign of $\bfp_L$, see Fig. \ref{fig-interchange} for an illustration of the modifications of the electronic upon changing the sign of $\bfp_L$. The interchange of the level broadenings does have a direct consequence on the total charge current, as will be discussed in the following.

First, consider conditions in which there is no molecular spin anisotropy, that is, $\bfepsilon_1=0$. Then, $G_0^r$ is even with respect to the sign of $\bfp_L$, while $\bfG_1^r$ is odd, which means that the induced spin-polarization in the molecule changes sign with the sign change of $\bfp_L$. Moreover, under those conditions, the level broadenings $1/\tau_\pm$ are degenerate, since $\tan\varphi=0$.
The direct consequence of these properties is that $\calT_{00}(-\bfp_L)=\calT_{00}(\bfp_L)$, and $\calT_{01}(-\bfp_L)\sim-\bfp_L\cdot(-A_0\bfB_1-\bfA_1B_0)$ and $\calT_{01}(-\bfp_L)\sim-\bfp_L\cdot(-A_0\bfB_1-\bfA_1B_0)$, such that the \emph{charge magneto-transmission} $\calT_0(\bfp_L)-\calT_0(-\bfp_L)$ vanishes identically. Hence, since the total charge current is also even with respect to the sign of $\bfp_L$, this entails that $J(\bfp_L)-J(-\bfp_L)=0$, that is, there is \emph{no} spin selectivity effect. By contrast, since $\calT_{\sigma\sigma'}\neq0$ and both $\calT_{10}$ and $\calT_{11}$ are odd with respect to $\bfp_L$, the difference $\calT_{\sigma\sigma'}(\up)-\calT_{\sigma\sigma'}(\down)$ is non-vanishing. Hence, despite the vanishing magneto-current, the spin-current is non-vanishing, since $\calT_1\neq0$. Also, the  \emph{spin magneto-transmission} $\calT_1(\bfp_L)-\calT_1(-\bfp_L)\neq0$. This example illustrates the inappropriateness of considering the spin-resolved transmission in the context of magneto-current studies, a conclusion that was also previously drawn in Ref. \citenum{JPhysChemC.125.23364}.

\begin{figure}[t]
\begin{center}
\includegraphics[width=\columnwidth]{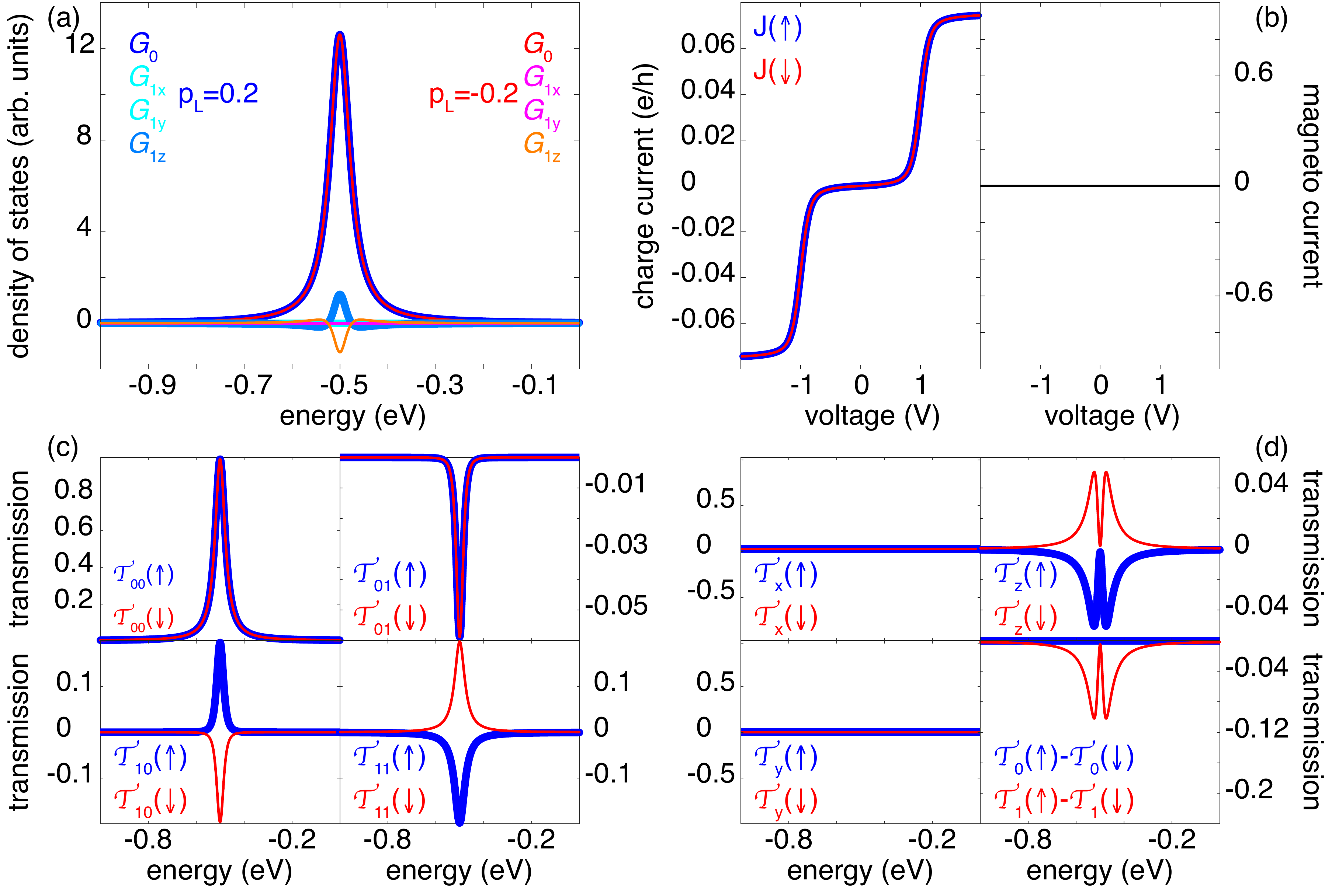}
\end{center}
\caption{Single level model with vanishing spin anisotropy, $\bfepsilon_1=0$ for $\bfp_L=0.2\hat{\bf z}$ (bold -- blue color scale) and $\bfp_L=-0.2\hat{\bf z}$ (faint -- red color scale).
(a) Charge, $-\im G_0^r(\omega)/\pi$, and spin, $-\im\bfG_1^r(\omega)/\pi$, resolved density of electron states.
(b) Charge current (left) and normalized magneto-current (right).
(c) Transmission coefficients. In the lower panels, the $x$- and $y$-components are zero (not shown).
(d) Spin-resolved transmissions and magneto-transmissions. In the lower right panel, the $x$- and $y$-components of $\calT_1(\up)-\calT_1(\down)$ are zero (not shown).
Parameters used are: $\dote{0}-\mu=-0.5$ eV, $\Gamma_0=0.1$ eV, and $T=300$ K.
}
\label{fig-SingleLevel}
\end{figure}

These observations are visualized in Figure \ref{fig-SingleLevel}, showing (a) the charge, $-\im G_0^r(\omega)/\pi$, and spin, $-\im\bfG_1^r(\omega)/\pi$, densities of states for $\bfp_L=0.2\hat{\bf z}$ (bold -- blue color scale) and $\bfp_L=-0.2\hat{\bf z}$ (faint -- red color scale), (b) corresponding charge current (left) and normalized magneto-current, $\tilde\Delta J=100\cdot\Delta J/\sum_\sigma J(\sigma)$, (right), (c) transmissions, and (d) spin-resolved transmissions and magneto-transmissions. The spin-polarization $\bfp_L$ in the left lead generates a finite spin-density in the longitudinal ($\hat{\bf z})$ direction, Figure \ref{fig-SingleLevel} (a), which changes polarity as $\bfp_L\rightarrow-\bfp_L$. This property is carried over to the transmissions $\calT_{mn}$, $m,n\in\{0,1\}$, see Figure \ref{fig-SingleLevel} (c), illustrating the even (odd) property of $\calT_{0m}$ ($\calT_{1m}$) as $\bfp_L\rightarrow-\bfp_L$. Consequently, the spin-resolved transmission as well as spin magneto-transmission are non-vanishing, Figure \ref{fig-SingleLevel} (d), while the charge magneto-transmission vanishes.

Second, should there be a non-vanishing molecular spin anisotropy, that is, $\bfepsilon_1\neq0$, the over all picture changes.

First it may be noticed that the two components $\calT_{00}$ and $\calT_{01}$ can be written as
\begin{subequations}
\begin{align}
\calT_{00}=&
	\biggl(\frac{\Gamma_0}{4}\biggr)^2
		\frac{(\omega-\dote{0})^2+\dote{1}^2+(\Gamma_0/4)^2(1+(p_L/2)^2)}
		{|(\omega-E_+)(\omega-E_-)|^2}
	,
\\
\calT_{01}=&
	\biggl(\frac{\Gamma_0}{4}\biggr)^2
	\bfp_L\cdot
		\frac{2\bfepsilon_1(\omega-\dote{0})-\bfp_L(\Gamma_0/4)^2}
		{|(\omega-E_+)(\omega-E_-)|^2}
	,
\end{align}
\end{subequations}
since $\bfA_1\times\bfB_1\sim\bfp_L\cdot(\bfepsilon_1\times\bfp_L)=0$. 
By contrast to conditions where $\bfepsilon_1=0$, the component $\calT_{01}$ is neither even nor odd with respect to the sign of $\bfp_L$, which is a direct consequence of the symmetry rules of the poles, see Eq. \eqref{eq-Esymmetries}. Hence, the difference $\calT_{01}(\bfp_L)-\calT_{01}(-\bfp_L)$ is non-vanishing, such that also the magneto-current $\Delta J\neq0$.

\begin{figure}[t]
\begin{center}
\includegraphics[width=\columnwidth]{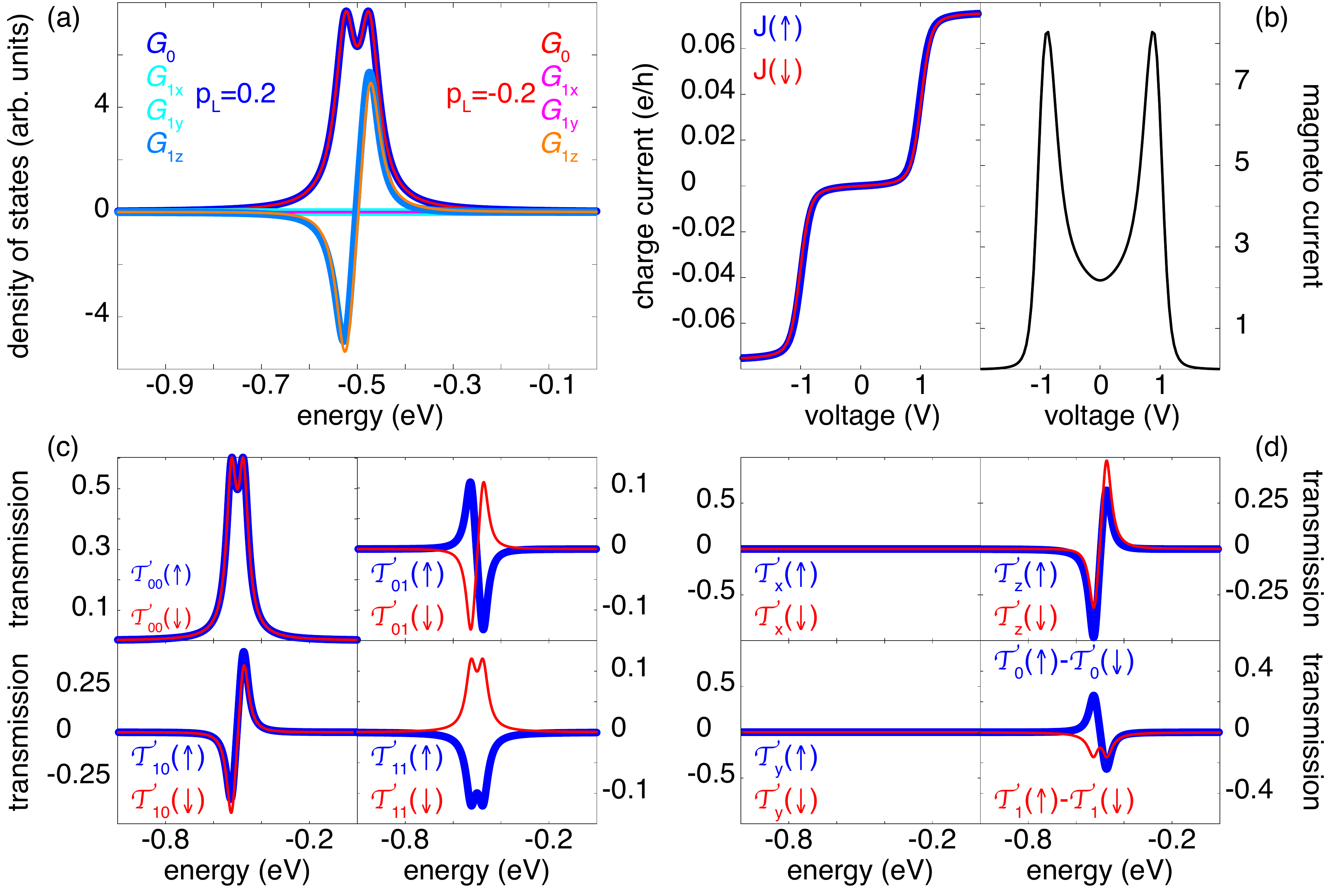}
\end{center}
\caption{Single level model with vanishing spin anisotropy, $\bfepsilon_1=\dote{1}\hat{\bf z}$, $\dote{1}=25$ meV, for $\bfp_L=0.2\hat{\bf z}$ (bold -- blue color scale) and $\bfp_L=-0.2\hat{\bf z}$ (faint -- red color scale).
(a) Charge, $-\im G_0^r(\omega)/\pi$, and spin, $-\im\bfG_1^r(\omega)/\pi$, resolved density of electron states.
(b) Charge current (left) and normalized magneto-current (right).
(c) Transmission coefficients. In the lower panels, the $x$- and $y$-components are zero (not shown).
(d) Spin-resolved transmissions and magneto-transmissions. In the lower right panel, the $x$- and $y$-components of $\calT_1(\up)-\calT_1(\down)$ are zero (not shown).
Other parameters are as in Figure \ref{fig-SingleLevel}.
}
\label{fig-SingleLevel001}
\end{figure}

The plots in Figure \ref{fig-SingleLevel001} illustrate (a) the charge and spin densities, (b) charge and normalized magneto-current, (c) transmissions, and (d) spin-resolved transmission and magneto-transmission, for the same set-up as in the previous example, however, with $\bfepsilon_1=25\hat{\bf z}$ meV. The non-vanishing intrinsic local spin moment increases the separation between the spin states, which is signified by the double peak structure in the charge density. It also induces an asymmetry in the spin density, with respect to the bare level energy $\dote{0}$. However, it can be seen that the spin density is neither even nor odd upon sign change of $\bfp_L$, which leads to a finite magneto-current, see Figure \ref{fig-SingleLevel001} (b). The corresponding transmissions, Figure \ref{fig-SingleLevel001} (c), illustrate $\calT_{01}$ and $\calT_{00}$, showing that the sum of their contributions is unchanged. It can also be noticed that the $z$-components of the transmissions $\calT_{1m}$ pertaining to the spin-current carry their properties over to the spin-resolved transmissions and magneto-transmissions, see Figure \ref{fig-SingleLevel001} (d). It should also be noticed, however, that while the magneto-transmission for the charge current is asymmetric with respect to the bare energy level $\dote{0}$, the corresponding magneto-transmission for the spin-current is symmetric. Hence, also in this situation where there is an intrinsic molecular spin anisotropy, the spin-resolved transmissions provide inappropriate information about the magneto-current.

Among the conclusions that can be drawn from these two examples, first one should notice that the spin-resolved transmission has little, if anything, to do with the mechanisms that are relevant for chiral induced spin selectivity since it is a measure for the spin current whereas chiral induced spin selectivity is measured through the magneto-current, which is the difference between two charge currents obtained in two separate measurements. Second, a non-vanishing magneto-current can only be provided in a system comprising a component, here, the molecule, which has an intrinsic spin anisotropy. For instance, the magneto-current should not originate in the spin-polarized leads, since these are included for supplying a reference spin-polarization of the injected current. Hence, such leads merely play a role for the detection of the effect. Third, while an effective single-electron theory can be used to capture the phenomenon of a magneto-current, one has to include a spin anisotropy in the description which, at this simple level of description, provides a phenomenological addition for which the microscopic origin may be unknown. Fourth, the two examples illustrate that despite the transmission itself may provide sufficient information about the transport properties to predict and interpret the experimental results, it is necessary to acquire this information for all configurations considered in the experiment. Only considering the transmission without the influence of, e.g., a ferromagnetic lead or not comparing the results for the ferromagnetic lead with two opposite spin-polarizations does not allow for a full comprehension of the phenomenon.

Finally, before ending this section an issue that has been raised within the community about the properties of the spin anisotropy is addressed. It has been conjectured that the spin anisotropy $\bfepsilon_1$ has to include a longitudinal component for the chiral induced spin selectivity effect to arise. This conjecture is, however, likely to have been made on a loose basis without a solid foundation.

Within the introduced example, it is easy to see that, e.g., a spin anisotropy $\bfepsilon_1=\dote{1}(\hat{\bf x}+\hat{\bf y})$ and spin-polarization $\bfp_L=p_L\hat{\bf z}$ certainly leads to that the product $\bfepsilon_1\cdot\bfp_L=0$,  and that both transmissions $\calT_{0m}$ are even with the sign of $\bfp_L$. However, for a lead which is also spin-polarized in the transverse direction, that is, $\bfp_L=\bfp_\perp+p_z\hat{\bf z}$, where $\bfp_\perp=|(p_x,p_y)|$ and at least one of $p_x$ and $p_y$ non-zero, the transmission $\calT_{01}$ loses its even property with the sign of $\bfp_L$.

\begin{figure}[t]
\begin{center}
\includegraphics[width=\columnwidth]{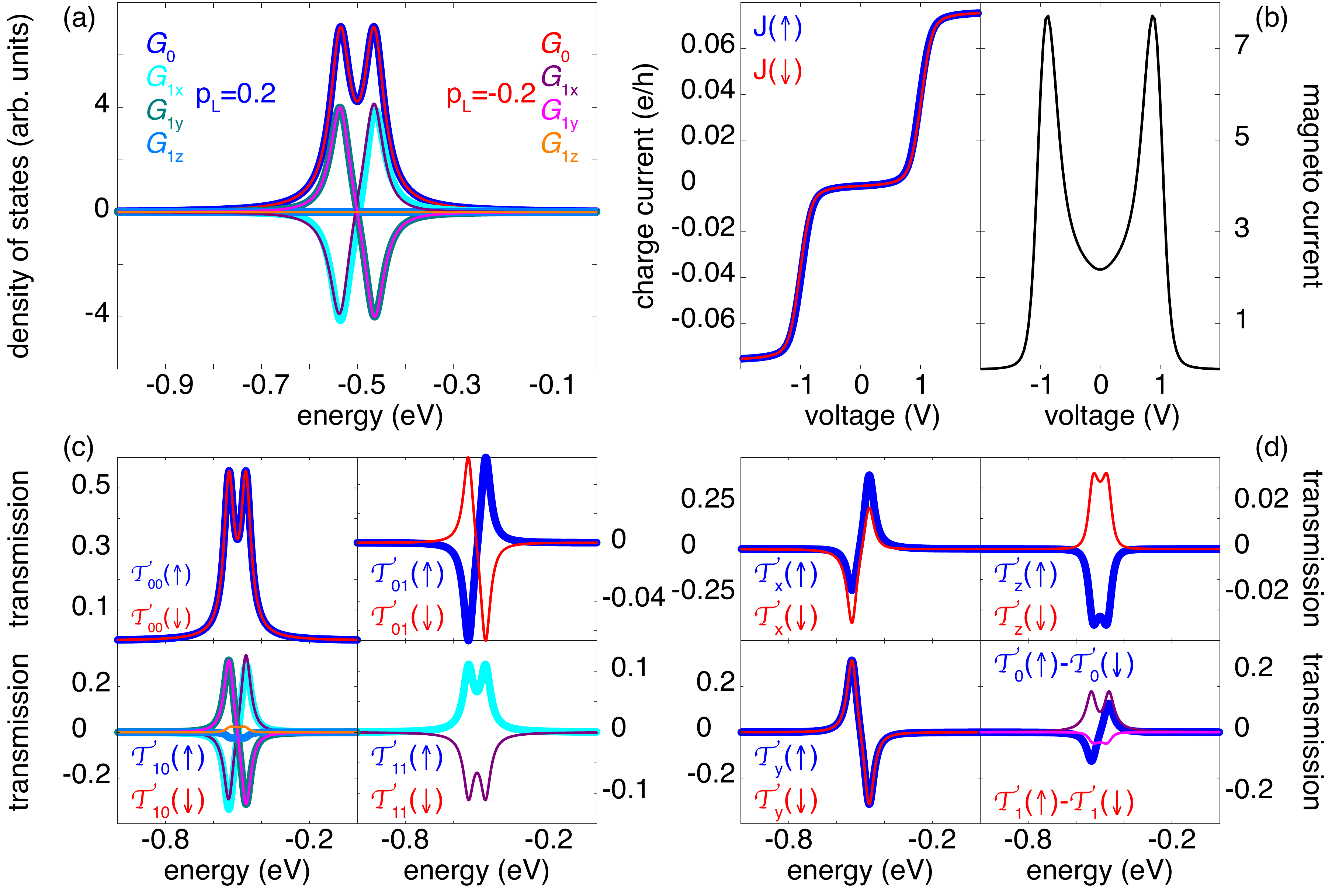}
\end{center}
\caption{Single level model with vanishing spin anisotropy, $\bfepsilon_1=\dote{1}(\hat{\bf x}+\hat{\bf y})$, $\dote{1}=25$ meV, for $\bfp_L=0.2\hat{\bf x}$ (bold -- blue color scale) and $\bfp_L=-0.2\hat{\bf x}$ (faint -- red color scale).
(a) Charge, $-\im G_0^r(\omega)/\pi$, and spin, $-\im\bfG_1^r(\omega)/\pi$, resolved density of electron states.
(b) Charge current (left) and normalized magneto-current (right).
(c) Transmission coefficients. In the lower left, color code is the same as in panel (a). In the lower right, the $y$- and $z$-components are zero (not shown).
(d) Spin-resolved transmissions and magneto-transmissions. In the lower right, color code for $\calT_1(\up)-\calT_1(\down)$ is the same as in panel (a).
Other parameters are as in Figure \ref{fig-SingleLevel}.
}
\label{fig-SingleLevel110}
\end{figure}

In Figure \ref{fig-SingleLevel110}, the results for an example with $\bfepsilon_1=\dote{1}(\hat{\bf x}+\hat{\bf y})$, $\dote{1}=25$ meV, and $\bfp_L=\pm0.2\hat{\bf x}$ are summarized, illustrating the non-vanishing transverse spin-polarization, (a), and corresponding magneto-current (b), as well as the transmission (c), spin-resolved transmission, and magneto-transmission (d). It is interesting to notice that despite the transverse characters of the spin-polarization and spin-anisotropy, there is, nonetheless, a non-vanishing longitudinal component to the spin magneto-transmission, see Figure \ref{fig-SingleLevel110} (d).

\begin{figure}[t]
\begin{center}
\includegraphics[width=\columnwidth]{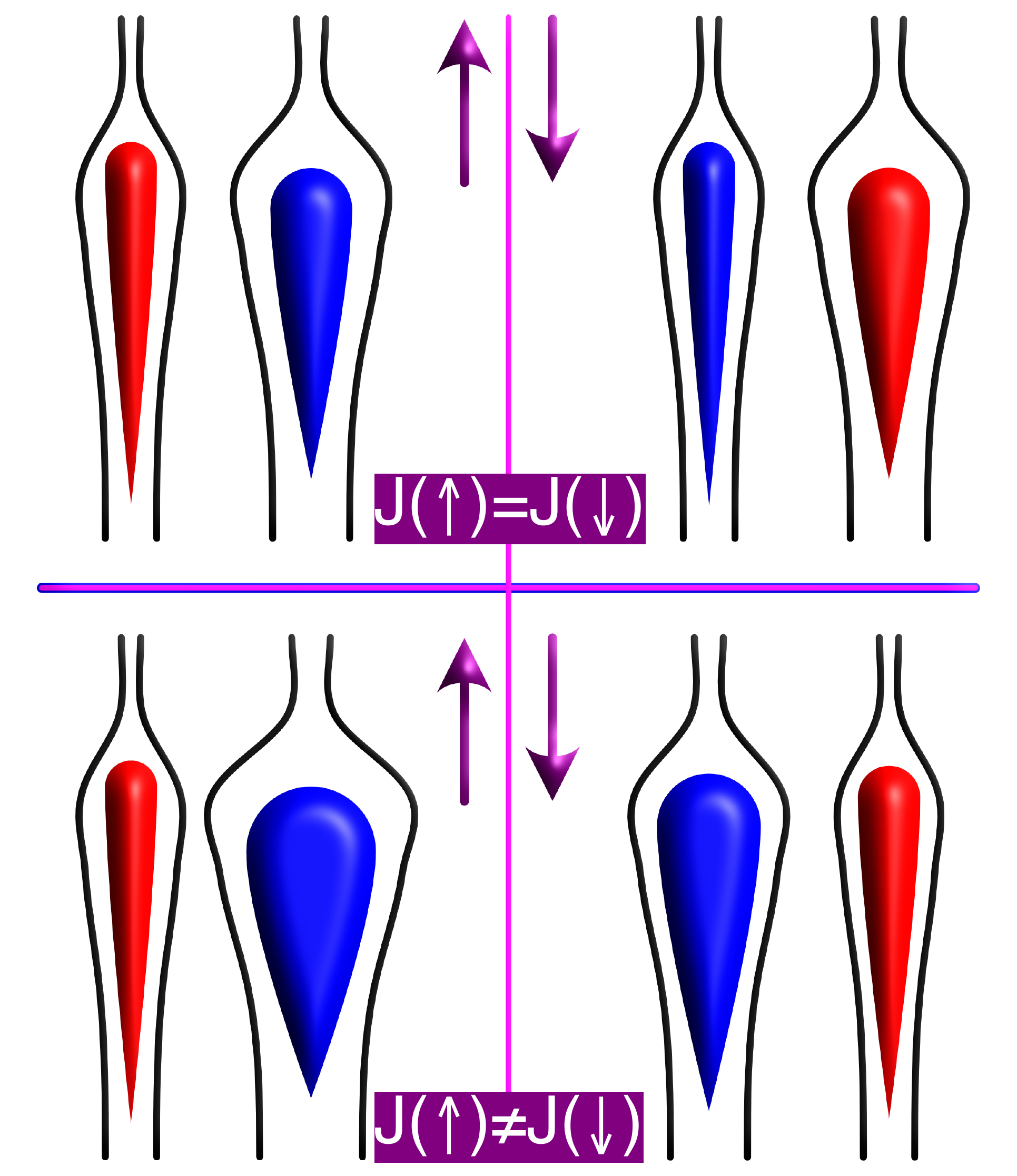}
\end{center}
\caption{Schematic illustration of how a single particle theory (upper) versus a correlated theory (lower) accounts for the chiral induced spin selectivity effect. The arrows represent the two different configurations of the external force fields, the red and blue objects represent the chiral molecule, whereas the black lines indicate the particle flux.}
\label{fig-schematicCISS}
\end{figure}

The conclusions of this discussion can be summarized in the following terms. First of all, in the transport set-up, the chiral induced spin selectivity effect refers to a measure of the charge currents obtained in one system under two very different conditions, or, configurations, which is outlined in Fig. \ref{fig-schematic}. This picture can be generalized into some intensities $I(m)$ obtained from the same system configured under conditions, say, $m=\up$ and $m=\down$. The two intensities are compared through the normalized difference $\Delta I=[I(\up)-I(\down)]/[I(\up)+I(\down)]$, and should it be non-vanishing there is a measurable difference between the two configurations. As it has been made clear in this discussion, the chiral induced spin selectivity effect is \emph{neither} a measure of the spin-polarized current \emph{nor} is it a measure of the spin-current through the system. Hence, the transmission is a quantity which very poorly accounts for the effect, first and foremost since the effect is a result of integrated degrees of freedom whereas the transmission relates to spectral properties.

Second, the chiral induced spin selectivity effect cannot be captured in a single particle theory unless an effective spin anisotropy is included, an anisotropy that inevitable originates from the interplay between structure, spin-orbit interactions, and electron correlations. The latter picture should, therefore, be referred to as a correlated description. The difference between the descriptions is summarized by means of the schematic illustration in Fig. \ref{fig-schematicCISS}. Here, the upper and lower panels represents transport properties using a single particle and correlated description, respectively. In the single particle description, the spin properties of the molecule are governed solely by the external force field, hence, the spin-polarizations of the molecule are equally strong but with opposite polarity in the two configurations. Hence, the total flux through the system is the same in the two configurations, although the spin-polarizations of the currents are different. By contrast, in the correlated description, there is an internal anisotropy which can either enhance or diminish the effect of the external force field. One may think of this as a coercive field that has to be overcome before the spin-polarization of the molecule aligns with the external force field. Hence, the resulting flux through the system is \emph{not} equal in the two configurations, since the spin anisotropy acts as an additional and \emph{variable} resistance to the charge current between the two configurations.


\subsection{Chiral lattice model}
Next, the theoretical basis discussed in the previous section is here transferred into an effective model for chiral molecules. For the purpose of being concrete, this model is based on the lattice model introduced in Refs. \citenum{JPhysChemLett.10.7126,PhysRevB.102.235416,NanoLett.21.3026,JPhysChemLett.13.808} with the modifications, however, that the single electron level per site is extended to include multiple levels. Furthermore, the interactions (electron-electron or electron-vibration) that were included in those previous discussions are here replaced by a simple phenomenological spin anisotropy.
The generic molecular geometry considered here is described by the set $\mathbb{M}=M\times N$ of spatial coordinates $\bfr_m=(a\cos\varphi_m,a\sin\varphi_m,c_m)$, $\varphi_m=2\pi(m-1)/(\mathbb{M}-1)$, $m=1,\ldots,\mathbb{M}$, and $c_m=c\varphi_m/2\pi$, where $a$ and $c$ define the radius and length, respectively, of the helical structure, whereas $M$ and $N$ denote the number of turns and ions per turn. Each coordinate denotes an ionic site which is represented by a set of electron levels described by
\begin{align}
\Hamil_m=&
	\sum_n\psi^\dagger_{mn}\bfepsilon_{mn}\psi_{mn}
	+
	\sum_{nn'}\Bigl(\psi^\dagger_{mn}\bfw_{mnn'}\psi_{mn'}+H.c.\Bigr)
	,
\label{eq-MLsite}
\end{align}
where $\psi_{mn}^\dagger=(\ddagger{mn\up}\ \ddagger{mn\down})$ ($\psi_{mn}$) is the creation (annihilation) spinor, $\bfepsilon_{mn}=\dote{mn}^{(0)}\sigma^0+\bfepsilon^{(1)}_{mn}\cdot\bfsigma$ represents the bare energy spectrum for electrons in the $n$th level at the site, and $\bfw_{mnn'}$ denotes the matrix element for transitions between energy levels within the site.

Electrons hopping between nearest-neighboring and next nearest-neighboring sites occur with the energies $\bft_{mmn'}$ and $i\lambda_{mmn'}\bfv_m^{(s)}\cdot\bfsigma$, and spin-orbit coupling is picked up between next-nearest neighbor sites through processes of the type $\psi_{mn}^\dagger\bft_{mnn'}\psi_{(m\pm s)n'}$ and $i\lambda_{mnn'}\psi_{mn}^\dagger\bfv_m^{(s)}\cdot\bfsigma\psi_{(m+2s)n'}$, respectively, where $s=\pm1$. Here, $\lambda_{mnn'}$ denotes the spin-orbit interaction parameter, whereas the vector $\bfv_m^{(s)}=\hat\bfd_{m+s}\times\hat\bfd_{m+2s}$ defines the chirality of the helical molecule in terms of the unit vectors $\hat{\bfd}_{m+s}=(\bfr_m-\bfr_{m+s})/|\bfr_m-\bfr_{m+s}|$; positive chirality corresponds to right handed helicity.

The chiral molecule comprising $\mathbb{M}$ sites can, thus, be modeled by the Hamiltonian
\begin{align}
\Hamil_\text{mol}=&
	\sum_{m=1}^{\mathbb{M}}
		\Hamil_m
	-
	\sum_{m=1}^{\mathbb{M}-1}
		\Bigl(
			\psi^\dagger_{mn}\bft_{mnn'}\psi_{(m+1)n'}
			+
			H.c.
		\Bigr)
\nonumber\\&
	+
	\sum_{m=1}^{\mathbb{M}-2}
		\Bigl(
			i\lambda_{mnn'}\psi^\dagger_{mn}\bfv_m^{(+)}\cdot\bfsigma\psi_{(m+2)n'}
			+
			H.c.
		\Bigr)
	.
\label{eq-MLmol}
\end{align}
The molecule is coupled to metallic leads by tunneling interactions
\begin{align}
\Hamil_T=&
	\sum_\bfp t_\bfp\psi^\dagger_\bfp\psi_1+\sum_\bfq t_\bfq\psi^\dagger_\bfq\psi_N+H.c.
	,
\end{align}
where the leads are modeled by $\Hamil_\chi=\sum_{\bfk\in\chi}\leade{\bfk}\psi^\dagger_\bfk\psi_\bfk$, $\leade{\bfk}=\dote{\bfk}+\Delta_\chi/2$. Here, $\chi=L,R$ denotes the left ($L$, $\bfk=\bfp$) or right ($R$, $\bfk=\bfq$) lead, and $\Delta_\chi$ denotes the spin-gap in the lead. Hence, the full metal-molecule-metal junction is described by
\begin{align}
\Hamil=&
	\Hamil_L+\Hamil_R+\Hamil_\text{mol}+\Hamil_T
	.
\label{eq-MLHamil}
\end{align}
The current is calculated using the formula in Eq. \eqref{eq-J}, which can be obtained by employing standard methods for non-equilibrium Green function, see, e.g., Refs. \citenum{JPhysChemLett.10.7126,PhysRevB.102.235416}.

\subsubsection{Single electron per site}

\begin{figure}[t]
\begin{center}
\includegraphics[width=\columnwidth]{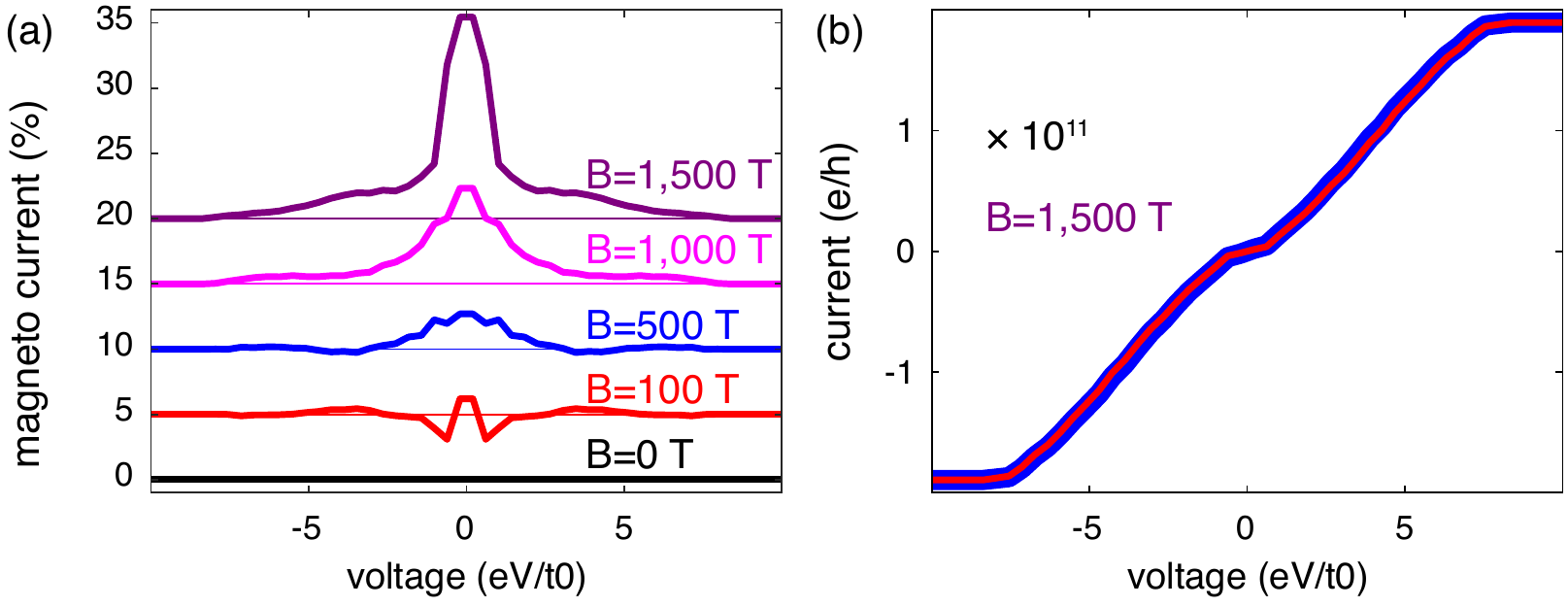}
\end{center}
\caption{
(a) Normalized magneto-current for a $10\times6$ sites helix as function of the voltage bias for different spin-gaps $\bfepsilon^{(1)}=g\mu_BB$, and $B=0$ (black),  $B=100$ T (red),  $B=500$ T (blue),  $B=1,000$ T (magenta), and $B=1,500$ T (purple). Notice that the curves are off-set for clarity.
(b) Charge currents for the set-up with $B=1,500$ T.
Parameters used are: $\dote{0}-\mu=-1/2$, $\lambda_0=1/1,000$, $\Gamma_0=1/10$ in units of $t_0=1$ eV, and $\bfp_L=\pm0.2\hat{\bf z}$ at $T=300$ K.
}
\label{fig-CISS10x6B}
\end{figure}

First, consider a single stranded helix, comprising a single electron level per site, that is, $n=1$ for each $m$. Such a system can be represented in terms of the model introduced in Eqs. \eqref{eq-MLsite}--\eqref{eq-MLHamil} by setting the intra-site hybridization matrix $\bfw_{mnn'}=0$, while putting the nearest neighbor and next-nearest neighbor tunneling matrices $\bft_{mnn'}=t_0\sigma^0$ and $\lambda_{mnn'}\bfv_m^{(s)}\cdot\bfsigma=\lambda_0\bfv_m^{(s)}\cdot\bfsigma$, assuming uniform tunneling rates throughout the structure. The last assumption is not necessary in the context of single stranded helices, however, it is implemented here for simplicity. For a chain of equivalent sites, it is, furthermore, justified to assume that $\bfepsilon_m=\bfepsilon=\dote{0}\sigma^0+\bfepsilon_1\cdot\bfsigma$. For the sake of illustrating the effect, it is instructive to express the internal spin-anisotropy in terms of an equivalent magnetic field $B$, such that $\bfepsilon_1=g\mu_BB$, where $g=2$ is the gyromagnetic ratio and $\mu_B$ is the Bohr magneton. Then, using $t_0=1$ eV, and setting $\lambda_0=t_0/1,000$, $\Gamma_0=t_0/10$, $\bfp_L=\pm0.2\hat{\bf z}$, and $\dote{0}-\mu=-t_0/2$, where $\mu$ is the common equilibrium chemical potential of the system, and performing the simulations at $T=300$ K, the resulting normalized magneto-current is shown in Figure \ref{fig-CISS10x6B} (a) for $B=0$ (black), $B=100$ T (blue), $B=500$ T (red), $B=1,000$ T (magenta), and $B=1,500$ T (purple). The currents for $\bfp_L=0.2\hat{\bf z}$ (blue) and $\bfp_L=-0.2\hat{\bf z}$ (red) in Figure \ref{fig-CISS10x6B} (b), correspond to the case $B=1,500$ T.

\begin{figure}[t]
\begin{center}
\includegraphics[width=\columnwidth]{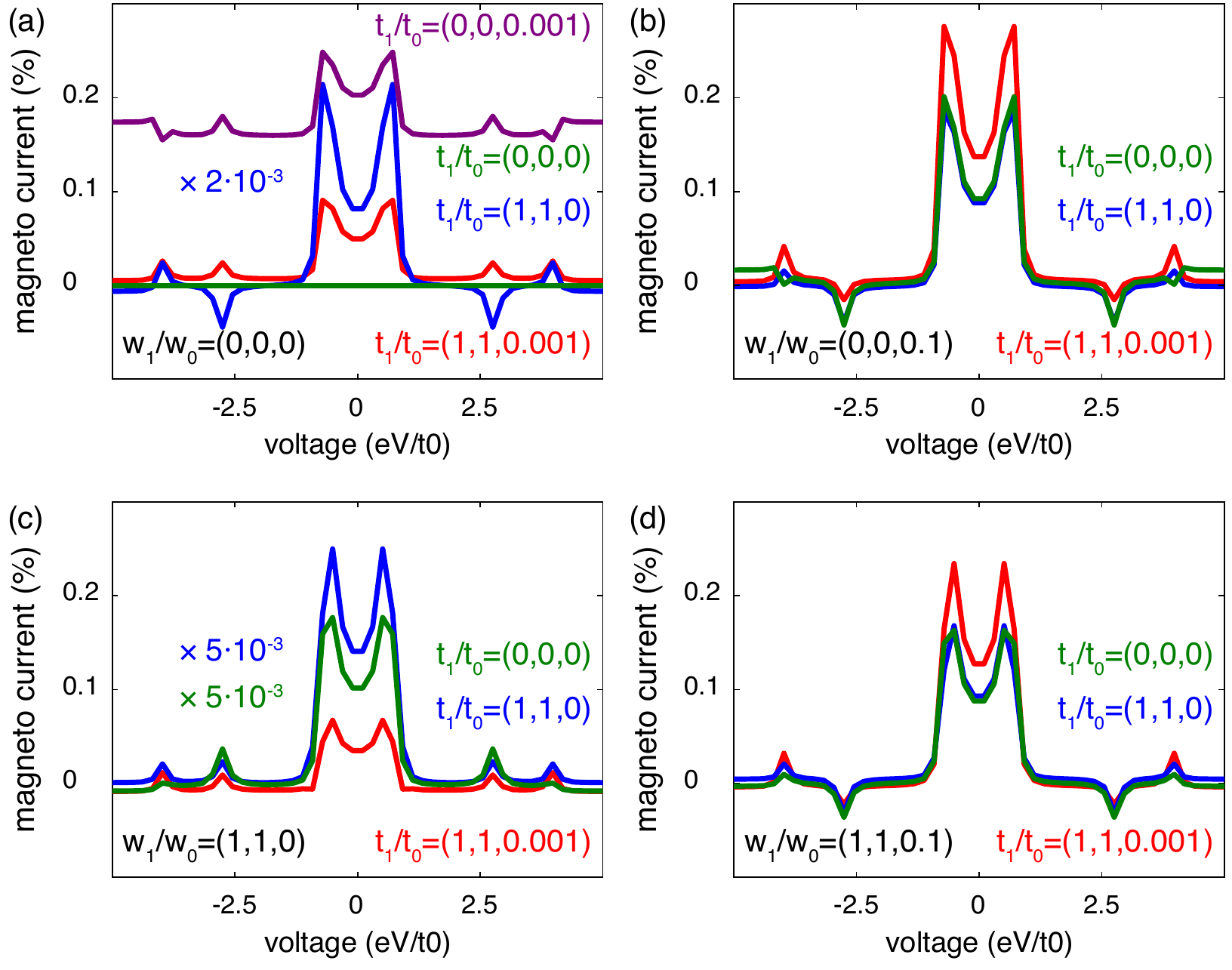}
\end{center}
\caption{
Normalized magneto-current for a $2\times3$ sites helix with two energy levels per site, for different intra-site and inter-site level hybridizations $\bfw_1$ and $\bft_1$, respectively.
(a) $\bfw_1/w_0=0$, 
(b) $\bfw_1/w_0=(0,0,1/10)$, 
(c) $\bfw_1/w_0=(1,1,0)$, and 
(d) $\bfw_1/w_0=(1,1,1/10)$,
where $\bft_1/t_0=(1,1,1/1,000)$ (red); $\bft_1/t_0=(1,1,0)$ (blue);  $\bft_1/t_0=0$ (green);  $\bft_1/t_0=(0,0,1/1,000)$ (purple).
Parameters used are: $\dote{0}-\mu=-1/2$, $\lambda_0=1/1,000$, $\Gamma_0=1/10$ in units of $t_0=1$ eV, and $\bfp_L=\pm0.2\hat{\bf z}$ at $T=300$ K.
}
\label{fig-CISS2x2x3deg}
\end{figure}

As can be expected, there is no magneto-current in absence of the spin anisotropy (black), while the it increases with increasing $B$. In the light of experimental observations where magneto-currents between a few percents to nearly a hundred percents, the values of the normalized magneto-currents plotted in Fig. \ref{fig-CISS10x6B} are reasonable. However, while the magnitudes of the equivalent $B$-fields are unrealistically large, they illustrate the inability to represent and explain the chiral induced spin selectivity effect in single stranded helices in terms of single electron theory using realistic values on the parameters. Despite this inability, the observation that the chiral induced spin selectivity effect actually can be obtained by inclusion of the spin anisotropy $\bfepsilon_1$, suggests that is should be possible to address the effect through electron correlations, as was previously done in, e.g., Refs. \citenum{JPhysChemLett.10.7126,PhysRevB.102.214303,PhysRevB.102.235416,NanoLett.21.3026,JACS.143.14235,JPhysChemC.125.23364,JPhysChemLett.13.808}.

\begin{figure}[t]
\begin{center}
\includegraphics[width=\columnwidth]{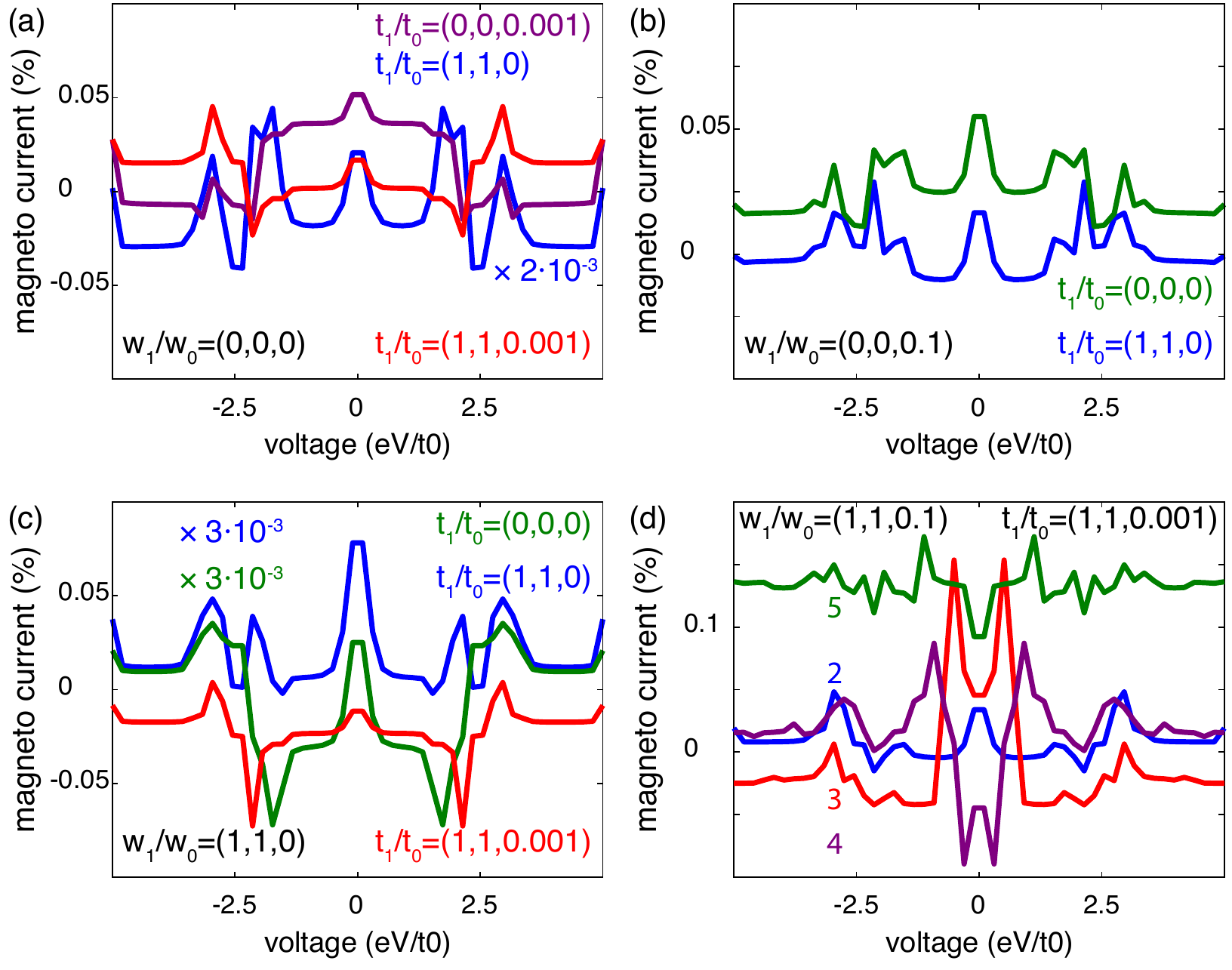}
\end{center}
\caption{
Normalized magneto-current for a $2\times3$ sites helix with two energy levels per site, for different intra-site and inter-site level hybridizations $\bfw_1$ and $\bft_1$, respectively.
(a) $\bfw_1/w_0=0$, 
(b) $\bfw_1/w_0=(0,0,0.1)$, 
(c) $\bfw_1/w_0=(1,1,0)$, and 
(d) $\bfw_1/w_0=(1,1,0.1)$,
where $\bft_1/t_0=(1,1,0.001)$ (red); $\bft_1/t_0=(1,1,0)$ (blue);  $\bft_1/t_0=0$ (green);  $\bft_1/t_0=(0,0,0.001)$ (purple).
Parameters used are: $\dote{0}-\mu=-0.5$, $\lambda_0=0.001$, $\Gamma_0=0.1t_0$ in units of $t_0=1$ eV, and $\bfepsilon_1=g\mu_BB$, $B=500$ T and $\bfp_L=\pm0.2\hat{\bf z}$ at $T=300$ K.
}
\label{fig-CISS2x2x3NonDeg}
\end{figure}

\subsubsection{Multiple electrons per site}

It has been argued that while the single stranded helix cannot be effectively modelled to provide a non-vanishing chiral induced spin selectivity effect \cite{PhysRevLett.108.218102,PNAS.11.11658,NewJPhys.22.113023}, this should be possible in a double stranded helix. Hence, whereas the example above can be regarded as to represent a single stranded helix, a double, or multiple, stranded helix can be represented in terms of the full model introduced in Eqs. \eqref{eq-MLsite}--\eqref{eq-MLHamil}, by letting the intra-site and inter-site hybridization matrices $\bfw_{mnn'}$, and $\bft_{mmn'}$, respectively, assume non-zero values, with $n=1,2,\ldots$ for each $m$. Hence, transitions from all $n$ to all $n'$ are allowed. For simplicity, however, it is assumed that $\bfw_{mnn'}=w_0\sigma^0+\bfw_1\cdot\bfsigma$, $\bft_{mnn'}=t_0\sigma^0+\bft_1\cdot\bfsigma$, and $\lambda_{mnn'}=\lambda_0$, that is, again assuming uniform hybridization rates throughout the structure. The values from the single stranded structure are retained, that is, $t_0=1$ eV, $\lambda_0=t_0/1,000$, and $\Gamma_0=t_0/10$.

Regarding the model to represent a double stranded helix with a single level per site, or a single stranded helix with two energy levels per site, it is first assumed that the energy levels are degenerate, such that $\bfepsilon_m=\dote{mn}^{(0)}\sigma^0$, and setting $\dote{mn}^{(0)}-\mu=\dote{0}-\mu=-t_0/2$, for all $m$ and $n$. In Fig. \ref{fig-CISS2x2x3deg}, the magneto-currents are plotted as function of the voltage bias for a $2\times3$ with two degenerate energy levels per $m$ with (a) $\bfw_1/w_0=(0,0,0)$, (b) $\bfw_1/w_0=(0,0,0.1)$, (c) $\bfw_1/w_0=(1,1,0)$, and (d) $\bfw_1/w_0=(1,1,0.1)$, and inter-site tunneling (green) $\bft_1/t_0=(0,0,0)$, (blue) $\bft_1/t_0=(1,1,0)$, (red) $\bft_1/t_0=(1,1,0.001)$, and (purple) $\bft_1/t_0=(0,0,0.001)$, and $w_0=-1/100$ in units of $t_0=1$ eV, while other parameters are as in Fig. \ref{fig-CISS10x6B}.

Under all conditions but one, there is a non-vanishing magneto-current, however, in all cases it is less than 0.3 \%. Absence of spin-dependent hybridization, that is, both $\bfw_{mnn'}=w_0\sigma^0$ and $\bft_{mnn'}=t_0\sigma^0$, yields a magneto-current which vanishes identically, see Fig. \ref{fig-CISS2x2x3deg} (a) (green). In all cases, the next-nearest neighbor spin-orbit coupling $\lambda_{mnn'}\neq0$. Hence, these results point towards an important aspect of the chiral structures. Namely, the mixing of the electronic structures of two separate helices, such that, at least one of $\bfw_{mnn'}^{(1)}$ and $\bft_{mnn'}^{(1)}$ are non-vanishing, introduces an intrinsic spin anisotropy in the structure which clearly is not there in the complete absence of this mixing. In this sense, then, these results corroborate the conclusions drawn in Refs. \citenum{PhysRevLett.108.218102,PNAS.11.11658,NewJPhys.22.113023}, that a double stranded structure is necessary to obtain the effect. However, these results also suggest that the magneto-current is likely to be less than a percent in a set-up with realistic values. The conclusions drawn in Refs. \citenum{PhysRevLett.108.218102,PNAS.11.11658,NewJPhys.22.113023} should be related to the inclusion of a dissipative contribution, something which would generate a large chiral induced spin selectivity effect, which also corroborates the conclusions from Refs. \citenum{JPhysChemLett.10.7126,PhysRevB.102.214303,PhysRevB.102.235416,NanoLett.21.3026,JACS.143.14235,JPhysChemC.125.23364,JPhysChemLett.13.808}.

It should, moreover, be noticed in the results presented in Fig. \ref{fig-CISS2x2x3deg}, that a sole transverse spin-dependent mixing $\bft_{mnn'}^{(1)}$ results in a non-vanishing magneto-current, see Fig. \ref{fig-CISS2x2x3deg} (blue). Despite this observation, it should also be noticed that the magneto-current is strongly suppressed in the absence of a longitudinal mixing. Hence, by introducing a longitudinal spin mixing component to either of $\bfw_{mnn'}^{(1)}$ or $\bft_{mnn'}^{(1)}$ enhances the magneto-current by about three orders of magntidue.

The picture remains qualitatively unchanged when lifting the degeneracy of the energy levels $\dote{mn}^{(0)}$, letting, for instance, $\dote{m1}^{(0)}<\dote{m2}^{(0)}<\ldots$. The simulated results for a $2\times3$ molecule with $\dote{m1}^{(0)}-\mu=-2t_0$ and $\dote{m2}^{(0)}-\mu=-t_0/2$ are summarized in Fig. \ref{fig-CISS2x2x3NonDeg} (a) $\bfw_1/w_0=(0,0,0)$, (b) $\bfw_1/w_0=(0,0,1/10)$, and (c) $\bfw_1/w_0=(1,1,0)$, with (green) $\bft_1/t_0=(0,0,0)$, (blue) $\bft_1/t_0=(1,1,0)$, (red) $\bft_1/t_0=(1,1,0.001)$, and (purple) $\bft_1/t_0=(0,0,0.001)$, and $w_0=-1/100$ in units of $t_0=1$ eV, while other parameters are as in Fig. \ref{fig-CISS10x6B}.

The lowered symmetry of the molecule is reflected in more details in the magneto-current, however, the overall amplitude of the magneto-current is reduced to less than 0.1 \%. This reduction in the amplitude can be understood as a reflection of the lowered symmetry, which leads to less resonant states in the structure and, hence, a weaker intrinsic spin anisotropy. Introducing more (non-degenerate) energy levels, equidistantly distributed in the energy range between $-2t_0$ and $-t_0/2$, Fig. \ref{fig-CISS2x2x3NonDeg} (d) (blue) two levels, (red) three levels, (purple) four levels, and (green) five levels, facilitates an enhancement of the magneto-current. This enhancement can be attributed to (i) the increasing number of conduction channels and (ii) the lowered spacing between adjacent energy levels, both conditions which increases the symmetry of the spectrum in the sense that more and more states become nearly resonant which, in turn, increases the intrinsic spin anisotropy.

In all the examples presented and discussed here, the amplitude of the normalized magneto-current is less than one percent, and it is worth to elaborate on reasons why this is the case. First, the chiral induced spin selectivity effect is not a measure of the induced spin-polarization in the molecule, or, in the composite system comprising the molecule and the leads, due to the spin-polarization of the injected electrons. The chiral induced spin selectivity effect is a measure of the variations of the charge current as function of the magnetic conditions applied to the system. The difference between the two concepts can be understood as the difference between the charge and spin currents, where only the former is directly related to the chiral induced spin selectivity effect.


\section{Conclusions}
\label{sec-conclusions}
In conclusion, it has been shown that the when comparing theoretical results with measurements of the chiral induced spin selectivity effect, one inevitably has to study the charge currents obtained for different configurations of the externally applied spin-polarization or magnetic field. The spin-polarized transmission of the system has little, if anything, to do with the conclusions one can draw about the effect for different compounds and configurations of the set-up. Likewise, while the spin-current may be used to extract deeper information about the origin of the chiral induced spin selectivity, thus far, there is no such experimental result with which the spin-current can be compared. It is, therefore, an important observation that studies of the chiral induced spin selectivity in terms of the spin-polarrized current or the spin current lead to conclusions that cannot be directly compared to any of the experimental observartions reported thus far. Reported observations of the chiral induced spin selectivity effect only concerns the properties that can be related to the charge currents. It has, furthermore, been demonstrated that while single-electron models, in general, are not viable tools to describe the chiral induces spin selectivity effect, by introducing an effective spin anisotropy to the molecular junction one can by, nonetheless, construct simple models as fitting tools. However, these effective spin anisotropies typically correspond to unrealistically large magnetic fields, which make the usefulness of such constructions questionable. Moreover, the origin of such spin anisotropies are related to the interplay between the structure, spin-orbit interactions, and electron correlations. The effective spin anisotropy can, in this sense, be thought of as a parametrization of the electron correlations. Finally, it was demonstrated that while the chiral induces spin selectivity effect can be enhanced by introducing more channels which are spin-dependently interconnected, the enhancement of the effect is still marginal, resulting in effects that are less than 1 \% in models using somewhat realistic parameter values.

This should conclude the discussion of the potential usefulness of single electron models for the description of the chiral induced spin selectivity effect, as well as to why the spin-resolved transmissions should not be used in comparison with the experimental observations of the effect.

\acknowledgements
The author thanks R. Naaman and Y. Dubi for constructive and fruitful discussions. Support from Vetenskapsr\aa det and Stiftelsen Olle Engkvist Byggm\"astare is acknowledged.



\begin{thebibliography}{200}
\bibitem{Science.283.814} Ray, K.; Ananthavel, S. P.; Waldeck, D. H.; Naaman, R., Asymmetric Scattering of Polarized Electrons by Organized Organic Films of Chiral Molecules, \emph{Science}, {\bf 1999}, \emph{283}, 814--816.
\bibitem{Science.331.894} G\"ohler, B.; Hamelbeck, V.; Markus, T. Z.; Kettner, M.; Hanne, G. F.; Vager, Z.; Naaman, R.; Zacharias, H., Spin Selectivity in Electron Transmission Through Self-Assembled Monolayers of Double-Stranded DNA, \emph{Science}, {\bf 2010}, \emph{331}, 894--897.
\bibitem{APLMaterials.9.040902} Waldeck, D. H.; Naaman, R.; Paltiel, Y., The spin selectivity effect in chiral materials, \emph{APL Materials}, {\bf 2021}, \emph{9}, 040902.


\bibitem{PNAS.110.14872} Mishra, D.; Markus, T. Z.; Naaman, R.; Kettner, M.; G\"ohler, B.; Zacharias, H.; Friedman, N.; Sheves, M.; Fontanesi, C., Spin-dependent electron transmission through bacteriorhodopsin embedded in purple membrane, \emph{Proc. Natl. Acad. Soc.}, {\bf 2013}, \emph{110}, 14872--14876.
\bibitem{NatComms.7.10744} Eckshtain-Levi, M.; Capua, E.; Refaely-Abramson, S.; Sarkar, S.; Gavrilov, Y.; Mathew, S. P.; Paltiel, Y.; Levy, Y.; Kronik, L.; Naaman, R., Cold denaturation induces inversion of dipole and spin transfer in chiral peptide monolayers, \emph{Nat. Comms.}, {\bf 2016}, \emph{7}, 10744.
\bibitem{AdvMat.30.1707390} Fontanesi, C.; Capua, E.; Paltiel, Y.; Waldeck, D. H.; Naaman, R., Spin-Dependent Processes Measured without a Permanent
Magnet, \emph{Adv. Mater.}, {\bf 2018}, \emph{30}, 1707390.
\bibitem{NanoLett.14.6042} Dor, O. B.; Morali, N.; Yochelis, S.; Baczewski, L. T.; Paltiel, Y., Local Light-Induced Magnetization Using Nanodots and Chiral
Molecules, \emph{Nano Lett.}, {\bf 2014}, \emph{14}, 6042--6049.
\bibitem{JPhysChemLett.9.2025} Kettner, M.; Maslyuk, V. V.; N\"urenberg, D.; Seibel, J.; Gutierrez, R.; Cuniberti, G.; Ernst, K. -H.; Zacharias, H., Chirality-Dependent Electron Spin Filtering by Molecular Monolayers of Helicenes, \emph{J. Phys. Chem. Lett.}, {\bf 2018}, \emph{9}, 2025--2030.
\bibitem{JPhysChemC.125.9875} Sang, Y.; Mishra, S.; Tassinari, F.; Karuppannan, S. K.; Carmieli, R.; Teo, R. D.; Migliore, A.; Beratan, D. N.; Gray, H. B.; Pecht, I.; Fransson, J.; Waldeck, D. H.; Naaman, R., Temperature Dependence of Charge and Spin Transfer in Azurin, \emph{J. Phys. Chem. C}, {\bf 2021}, \emph{125}, 9875--9883.
\bibitem{Chirality.33.93} M\"ollers, P. V.; Ulku, S.; Jayarathna, D.; Tassinari, F.; N\"urenberg, D.; Naaman, R.; Achim, C.; Zacharias, H., Spin-selective electron transmission through self-assembled monolayers of double-stranded peptide nucleic acid, \emph{Chirality}, {\bf 2021}, \emph{33}, 93.

\bibitem{NanoLett.11.4652} Xie, Z.; Markus, T. Z.; Cohen, S. R.; Vager, Z.; Gutierrez, R.; Naaman, R., Spin Specific Electron Conduction through DNA Oligomers, \emph{Nano Lett.}, {\bf 2011}, \emph{11}, 4652--4655.
\bibitem{JPhysChemLett.11.1550} Ghosh, S.; Mishra, S.; Avigad, E.; Bloom, B. P.; Baczewski, L. T.; Yochelis, S.; Paltiel, Y.; Naaman, R.; Waldeck, D. H., Effect of Chiral Molecules on the Electron’s Spin Wavefunction at Interfaces, \emph{J. Phys. Chem. Lett.}, {\bf 2020}, \emph{11}, 1550--1557.
\bibitem{AdvMater.28.1957} Kiran, V.; Mathew, S. P.; Cohen, S. R.; Delgado, I. H,; Lacour, J.; Naaman, R., Helicenes—A New Class of Organic Spin Filter, \emph{Adv. Mater}, {\bf 2016}, \emph{28}, 1957--1962.
\bibitem{NanoLett.19.5167} H. Alpern, K. Yavilberg, T. Dvir, N. Sukenik, M. Klang, S. Yochelis, H. Cohen, E. Grosfeld, H. Steinberg, Y. Paltiel, and O. Millo, Magnetic-related State and Order Parameter Induced in a Conventional Superconductor by Nonmagnetic Chiral Molecules, \emph{Nano Lett.}, {\bf 2019}, \emph{19}, 5167--5175.
\bibitem{ACSNano.14.16624} Mondal, A. K.; Brown, N.; Mishra, S.; Makam, P.; Wing, D.; Gilead, S.; Wiesenfeld, Y.; Leitus, G.; Shimon, L. J. W.; Carmieli, R.; Ehre, D.; Kamieniarz, G.; Fransson, J.; Hod, O.; Kronik, L.; Gazit, E.; Naaman, R., Long-Range Spin-Selective Transport in Chiral Metal-Organic Crystals with Temperature-Activated Magnetization, \emph{ACS Nano}, {\bf 2020}, \emph{14}, 16624--16633. 

\bibitem{JPhysChemLett.10.1139} Smolinsky, E. Z. B.; Neubauer, A.; Kumar, A.; Yochelis, S.; Capua, E.; Carmieli, R.; Paltiel, Y.; Naaman, R.; Michaeli, K., Electric Field-Controlled Magnetization in GaAs/AlGaAs Heterostructures-Chiral Organic Molecules Hybrids, \emph{J. Phys. Chem. Lett.}, {\bf 2019}, \emph{10}, 1139--1145.

\bibitem{NatComms.4.2256} Dor, O. B.; Yochelis, S.; Mathew, S. P.; Naaman, R.; Paltiel, Y., A chiral-based magnetic memory device without a permanent magnet, \emph{Nat. Comms.}, {\bf 2013}, \emph{4}, 2256.

\bibitem{NatComms.8.14567} Dor, O. B.; Yochelis, S.; Radko, A; Vankayala, K.; Capua, E.; Capua, A.; Yang, S. -H.; Baczewski, L. T.; Parkin, S. S. P.; Naaman, R.; Paltiel, Y., Magnetization switching in ferromagnets by adsorbed chiral molecules without current or external magnetic field, \emph{Nat. Comms.}, {\bf 2016}, \emph{8}, 14567.


\bibitem{PhysRevLett.124.166602} Inui, A.; Aoki, R.; Nishiue, Y.; Shiota, K.; Kousaka, Y.; Shishido, H.; Hirobe, D.; Suda, M.; Ohe, J.; Kishine, J.; Yamamoto, H. M.; Togawa, Y., Chirality-Induced Spin-Polarized State of a Chiral Crystal CrNb$_3$S$_6$, \emph{Phys. Rev. Lett.}, {\bf 2021}, \emph{124}, 166602.
\bibitem{PhysRevLett.127.126602} Shiota, K.; Inui, A.; Hosaka, Y.; Amano, R.; \={O}nuki, Y.; Hedo, M.; Nakama, T.; Hirobe. D; Ohe, J.; Kishine, J.; Yamamoto, H. M.; Shishido, H.; Togawa, Y., Chirality-Induced Spin Polarization over Macroscopic Distances in Chiral Disilicide Crystals, \emph{Phys. Rev. Lett.}, {\bf 2021}, \emph{127}, 126602.

\bibitem{JChemPhys.131.014707} Yeganeh, S.; Ratner, M. A.; Medina, E.; Mujica, V., Chiral electron transport: Scattering through helical potentials. \emph{J. Chem. Phys.}, {\bf 2009}, \emph{131}, 014707.
\bibitem{EPL.99.17006} Medina, E.; L\'opez, F.; Ratner, M. A.; Mujica, V., Chiral molecular films as electron polarizers and polarization modulators, \emph{EPL}, {\bf 2012}, \emph{99}, 17006.
\bibitem{JPCM.26.015008} Varela, S.; Medina, E.; L\'opez, F.; Mujica, V., Inelastic electron scattering from a helical potential: transverse polarization and the structure factor in the single scattering approximation, \emph{J. Phys.: Condens. Matter}, {\bf 2013}, \emph{26}, 015008.
\bibitem{PhysRevB.88.165409} Eremko, A. A.; Loktev, V. M.; Spin sensitive electron transmission through helical potentials, \emph{Phys. Rev. B}, {\bf 2013}, \emph{88}, 165409.
\bibitem{JChemPhys.142.194308} Medina, E.; Gonz\'alez-Arraga, L. A.; Finkelstein-Shapiro, D.; Berche, B.; Mujica, V., Continuum model for chiral induced spin selectivity in helical molecules, \emph{J. Chem. Phys.}, {\bf 2015}, \emph{142}, 194308.

\bibitem{PhysRevE.98.052221} D\'iaz, E.; Conteras, A.; Hern\'andez, J.; Dom\'inguez-Adame, F., Effective nonlinear model for electron transport in deformable helical molecules, \emph{Phys. Rev. E}, {\bf 2018}, \emph{98}, 052221.
\bibitem{PhysRevB.99.024418} Yang, X.; van der Wal, C. H.; van Wees, B. J.; Spin-dependent electron transmission model for chiral molecules in mesoscopic devices, \emph{Phys. Rev. B}, {\bf 2019}, \emph{99}, 024418.
\bibitem{NJP.20.043055} D\'iaz, E.; Albares, P.; Est\'evez, P. G.; Cerver\'o, J. M.; Gaul, C.; Diez, E.; Dom\'inguez-Adame, F., Spin dynamics in helical molecules with nonlinear interactions, \emph{New. J. Phys.}, {\bf 2018}, \emph{20}, 043055.
\bibitem{JPhysChemC.123.17043} Michaeli, K.; Naaman, R., Origin of Spin-Dependent Tunneling Through Chiral Molecules, \emph{J. Phys. Chem. C}, {\bf 2019}, \emph{123}, 17043--17048.

\bibitem{PhysRevB.85.081404(R)} Gutierrez, R; D\'iaz, E.; Naaman, R,; Cuniberti, G.; Spin-selective transport through helical molecular systems, \emph{Phys. Rev. B}, {\bf 2012}, \emph{85}, 081404(R).
\bibitem{PhysRevLett.108.218102} Guo, A. -M.; Sun, Q. -F., Spin-Selective Transport of Electrons in DNA Double Helix, \emph{Phys. Rev. Lett.}, {\bf 2012}, \emph{108}, 218102.
\bibitem{PNAS.11.11658} Guo, A. -M.; Sun, A. -F., Spin-dependent electron transport in protein-like single-helical molecules, \emph{Proc. Natl. Acad. Soc.}, {\bf 2014}, \emph{11}, 11658--11662.
\bibitem{JPhysChemC.117.13730} Rai, D.; Galperin, M., Electrically Driven Spin Currents in DNA, \emph{J. Phys. Chem. C}, {\bf 2013}, \emph{117}, 13730--13737.
\bibitem{PhysRevB.93.075407} Matityahu, S.; Utsumi, Y.; Aharony, A.; Entin-Wohlman, O.; Balseiro, C. A., Spin-dependent transport through a chiral molecule in the presence of spin-orbit interaction and nonunitary effects, \emph{Phys. Rev. B}, {\bf 2016}, \emph{93}, 075407.
\bibitem{PhysRevB.93.155436} Varela, S.; Mujica, V.; Medina, E., Effective spin-orbit couplings in an analytical tight-binding model of DNA: Spin filtering and chiral spin transport, \emph{Phys. Rev. B}, {\bf 2016}, \emph{93}, 155436.
\bibitem{ChemPhys.477.61} Behnia, S.; Fathizadeh, S.; Akhshani, A., Modeling spin selectivity in charge transfer across the DNA/Gold interface, \emph{Chem. Phys.}, {\bf 2016}, \emph{477}, 61--73.

\bibitem{JPhysChemLett.9.5453} Maslyuk, V. V.; Gutierrez, R.; Dianat, A.; Mujica, V.; Cuniberti, G., Enhanced Magnetoresistance in Chiral Molecular Junctions, \emph{J. Phys. Chem. Lett.}, {\bf 2018}, \emph{9}, 5453--5459.
\bibitem{JPhysChemLett.9.5753} D\'iaz, E.; Dom\'inguez-Adame, F.; Gutierrez, R.; Cuniberti, G.; Mujica, V., Thermal Decoherence and Disorder Effects on Chiral-Induced Spin Selectivity, \emph{J. Phys. Chem Lett}, {\bf 2018},\emph{9}, 5753--5459.
\bibitem{JChemTheoryComput.16.2914} Z\"ollner, M. S.; Varela, S.; Medina, E.; Mujica, V.; Herrmann, C., Insight into the Origin of Chiral-Induced Spin Selectivity from a Symmetry Analysis of Electronic Transmission, \emph{J. Chem. Theory Comput.}, {\bf 2020}, \emph{16}, 2914--2929.

\bibitem{CommunPhys.3.178} Ghazaryan, A.; Lemeshko, M.; Volosniev, A. G., Spin Filtering in Multiple Scattering off Point Magnets. \emph{Commun. Phys.}, {\bf 2020}, \emph{3}, 178.
\bibitem{NewJPhys.22.113023} Shitade, A.; Minamitani, E., Geometric Spin-Orbit Coupling and Chirality-Induced Spin Selectivity. \emph{New J. Phys.}, {\bf 2020}, \emph{22}, 113023.


\bibitem{PhysRevB.102.035431} Du, G. -H.; Fu, H. -H.; Wu, R., Vibration-enhanced spin-selective transport of electrons in the DNA double helix, \emph{Phys. Rev. B}, {\bf 2020}, \emph{102}, 035431.

\bibitem{NanoLett.21.6696} Fay, T. P.; Limmer, D. T., Origin of Chirality Induced Spin Selectivity in Photoinduced Electron Transfer, \emph{Nano Lett.}, {\bf 2021}, \emph{21}, 6696--6702.

\bibitem{JPhysChemLett.12.10262} Wang, C.; Guo, A. -M.; Sun, Q. -F.; Yan, Y., Efficient Spin-Dependent Charge Transmission and Improved Enantioselective Discrimination Capability in Self-Assembled Chiral Coordinated Monolayers, \emph{J. Phys. Chem. Lett.}, {\bf 2021}, \emph{12}, 10262--10269.

\bibitem{NanoLett.21.10423} Wang, C. Z.; Mujica, V.; Lai, Y. -C., Spin Fano Resonances in Chiral Molecules: An Alternative Mechanism for the CISS Effect and Experimental Implications, \emph{Nano Lett.}, {\bf 2021}, \emph{21}, 10423--10430.


\bibitem{JChemPhys.154.110901} Bian, X.; Yanze Wu, Y.; Teh, H. -H.; Zhou, Z.; Subotnik, J. E., Modeling nonadiabatic dynamics with degenerate electronic states, intersystem crossing, and spin separation: A key goal for chemical physics, \emph{J. Chem. Phys.}, {\bf 2021}, \emph{154}, 110901.
\bibitem{PhysRevB.104.024430} Volosniev. A. G.; Alpern, H.; Paltiel, Y.; Millo, O.; Lemeshko, M.; Ghazaryan, A., Interplay between friction and spin-orbit coupling as a source of spin polarization, \emph{Phys. Rev. B.}, {\bf 2021}, \emph{104}, 024430.
\bibitem{NanoLett.21.8190} Hoff, D. A.; Rego, L. G. C., Chirality-Induced Propagation Velocity Asymmetry, \emph{Nano Lett.}, {\bf 2021}, \emph{21}, 8190--8196.


\bibitem{NanoLett.19.5253} Dalum, S.; Hedeg\aa rd, P., Theory of Chiral Induced Spin Selectivity, \emph{Nano Lett.}, {\bf 2019} \emph{19}, 5253--5259.

\bibitem{JPhysChemLett.10.7126} Fransson, J., Chirality Induced Spin Selectivity: The Role of Electron Correlations, \emph{J. Phys. Chem. Lett.}, {\bf 2019}, \emph{10}, 7126--7132.

\bibitem{NanoLett.20.7077} Dianat, A.; Gutierrez, R.; Alpern, H.; Mujica, V.; Ziv, A.; Yochelis, S.; Millo, O.; Paltiel, Y.; Cuniberti, G., Role of Exchange Interactions in the Magnetic Response and Intermolecular Recognition of Chiral Molecules, \emph{Nano Lett.}, {\bf 2020}, \emph{20}, 7077--7086.

\bibitem{PhysRevB.102.214303} Zhang, L.; Hao, Y.; Qin, W,; Xie, S.; Qu, F., Chiral-induced spin selectivity: A polaron transport model, \emph{Phys. Rev. B}, {\bf 2020}, \emph{102}, 214303

\bibitem{PhysRevB.102.235416} Fransson, J., Vibrational origin of exchange splitting and chiral induced spin selectivity, \emph{Phys. Rev. B}, {\bf 2020}, \emph{102}, 235416.
\bibitem{NanoLett.21.3026} Fransson, J., Charge Redistribution and Spin Polarization Driven by Correlation Induced Electron Exchange in Chiral Molecules, \emph{Nano Lett.}, {\bf 2021}, \emph{21}, 3026--3032.

\bibitem{JACS.143.14235} Alwan, S.; Dubi, Y., Spinterface Origin for the Chirality-Induced Spin-Selectivity Effect, \emph{J. Am. Chem. Soc.}, {\bf 2021}, \emph{143}, 14235--14241.

\bibitem{JPhysChemC.125.23364} Huisman, K. H.; Thijssen, J. M., CISS Effect: A Magnetoresistance Through Inelastic Scattering, \emph{J. Phys. Chem. C}, {\bf 2021}, \emph{125}, 23364--23369.

\bibitem{JPhysChemLett.13.808} Fransson, J., Charge and Spin Dynamics and Enantioselectivity in Chiral Molecules, \emph{J. Phys. Chem. Lett} {\bf 2021}, \emph{13}, 808--814.

\bibitem{arXiv.2111.12917} Kato, A.; Yamamoto, H. M.; Kishine, J. -I., Chirality-Induced Spin Filtering in Pseudo Jahn-Teller Molecules, \emph{unpublished}, {\bf 2021}; arXiv.2111.12917.


\bibitem{PhysRevB.72.075314} Fransson, J., Nonequilibrium theory for a quantum dot with arbitrary on-site correlation strength coupled to leads, \emph{Phys. Rev. B}, {\bf 2005}, \emph{72}, 075314.

\bibitem{PhysRevB.50.5528} Jauho, A. -P.; Wingreen, N. S.; Meir, Y., Time-dependent transport in interacting and noninteracting resonant-tunneling systems, \emph{Phys. Rev. B.}, {\bf 1994}, \emph{50}, 5528--5544.

\bibitem{NonEquilibriumNanoPhysics} Fransson, J., \emph{Non-Equilibrium Nano-Physics}, Springer: Dordrecht, 2010.

\bibitem{PhysRevB.72.045415} Fransson, J., Angular conductance resonances of quantum dots strongly coupled to noncollinearly oriented ferromagnetic leads, \emph{Phys. Rev. B}, {\bf 2005}, \emph{72}, 045415.



\end{thebibliography}
\end{document}